%% file: main.tex
\documentclass[reprint, preprintnumbers, aps, prd, floatfix, superscriptaddress, nofootinbib]{revtex4-2}

\usepackage{silence}
\usepackage{amssymb, amsmath, physics, graphicx, xcolor, hyperref, mathtools, orcidlink, fontawesome, booktabs}

\newcommand{\bob}{\text{BOB}}
\newcommand{\res}{\text{res}}
\newcommand{\qnm}{\text{QNM}}

\newcommand{\iu}{\mathrm{i}\mkern1mu}
\newcommand{\du}{\mathrm{d}}
\newcommand{\lm}{\ell m}
\newcommand{\Ylm}{Y_{\ell m}}
\newcommand{\rstar}{r_{\ast}}

\newcommand{\IoA}{Institute of Astronomy, University of Cambridge, Madingley Road, Cambridge, CB3 0HA, UK}
\newcommand{\KICC}{Kavli Institute for Cosmology, University of Cambridge, Madingley Road, Cambridge, CB3 0HA, UK}
\newcommand{\DAMTP}{Department of Applied Mathematics and Theoretical Physics, Centre for Mathematical Sciences, University of Cambridge, Wilberforce Road, Cambridge, CB3 0WA, UK}
\newcommand{\wvu}{Department of Physics and Astronomy, West Virginia University, Morgantown, WV 26506, USA}
\newcommand{\centerGWcosmo}{Center for Gravitational Waves and Cosmology, West Virginia University, Chestnut Ridge Research Building, Morgantown, WV 26505, USA}

\begin{document}

\title{Moving Beyond Quasinormal Modes and Reinterpreting Direct Waves in Binary Black Hole Merger-Ringdowns}

\author{Richard Dyer\,\orcidlink{0009-0008-3720-6092}}
\email{richard.dyer@ast.cam.ac.uk}
\affiliation{\IoA}

\author{Anuj Kankani\,\orcidlink{0000-0002-7422-9137}}
\email{anuj.kankani@mail.wvu.edu}
\affiliation{\wvu} \affiliation{\centerGWcosmo}

\author{Gareth Arturo Marks \orcidlink{0009-0003-3160-9337}}
\email{gam54@cam.ac.uk}
\affiliation{\DAMTP}

\author{Sean T.\ McWilliams\,\orcidlink{0000-0003-2397-8290}}
\email{sean.mcwilliams@mail.wvu.edu}
\affiliation{\wvu} \affiliation{\centerGWcosmo}

\author{Christopher J.\ Moore\,\orcidlink{0000-0002-2527-0213}}
\email{christopher.moore@ast.cam.ac.uk}
\affiliation{\IoA} \affiliation{\DAMTP} \affiliation{\KICC}

\date{\today}

\begin{abstract}
    Based on black-hole perturbation-theory calculations and an effective-one-body picture of binary black hole mergers, it has been suggested that a \textit{direct wave} may be present and measurable in the ringdown. 
    Furthermore, it has been suggested that this signal might originate from substantially inside the light ring (or photon sphere) of the remnant black hole, perhaps even from close to the event horizon. 
    In this paper, we present three independent strands of evidence against such claims. 
    In the process, we show that there is indeed an early-time part of the ringdown signal that cannot be directly modeled by quasinormal modes.
    However, there is no evidence that this is a direct wave as it has previously been defined, and we show that the properties of this component are related to the remnant's light ring rather than its horizon.
\end{abstract}

\maketitle

\section{Introduction} \label{sec:intro}

The gravitational-wave (GW) signal from the merger and ringdown of binary black holes (BHs) carries information about the newly formed remnant object. 
The ringdown is dominated by quasinormal modes (QNMs) \cite{Vishveshwara:1970zz, Press:1971wr, 1999LRR.....2....2K, 1999CQGra..16R.159N, Berti:2005ys, 2009CQGra..26p3001B, 2026CQGra..43l3001B}: exponentially damped oscillations at well-defined eigenfrequencies of linear perturbations of the remnant BH. 
QNM-based models underpin many of the most constraining tests of the Kerr metric \cite{2019PhRvL.123k1102I, 2023PhRvL.131v1402C}, general relativity \cite{LIGOScientific:2019fpa,LIGOScientific:2020tif,LIGOScientific:2021sio, LIGOScientific:2026wpt}, and fundamental physics \cite{2021PhRvL.127a1103I, LIGOScientific:2025rid} using GWs.

The properties of QNMs are closely related to the location of the unstable circular null geodesic orbit known as the light ring \cite{Davis:1971gg, 1972ApJ...172L..95G, Mashhoon:1985cya} (or photon sphere), or, more precisely, to the peak of the effective potential that governs the propagation of linear waves around the BH.
For high-$\ell$ QNMs, this dependence is made explicit by the eikonal approximation for the QNM frequencies \cite{Press:1971wr, 1972ApJ...172L..95G, 1984PhRvD..30..295F, 2009PhRvD..79f4016C, 2009CQGra..26p3001B}, but the relationship is thought to hold qualitatively even for the low-$\ell$ QNMs.
This is consistent with the \textit{standard picture} in which the appearance of a BH to external observers is governed by its light ring (see, e.g.\ Ref.~\cite{2016PhRvL.116q1101C} for GWs and Ref.~\cite{Ames} for electromagnetic waves).
Without invoking contrived thought experiments, it is practically impossible to observe anything from significantly inside the light ring.
The light ring functions as a soft boundary on what our observations can probe, as opposed to the hard boundary provided by the event horizon.

However, QNMs are not the whole story. 
Mathematically, QNMs do not form a complete basis for describing the ringdown signal. 
Physically, there are other contributions to the ringdown. 
At early times, nonlinearities from the merger \cite{2020arXiv200400671O} are expected to be important; higher-order nonlinear QNMs \cite{Lagos:2022otp, Loutrel:2020wbw, Ripley:2020xby} (including quadratic \cite{2023PhRvL.130h1401C, 2023PhRvL.130h1402M} and cubic \cite{2026PhRvL.136s1403D} modes) are expected throughout the ringdown; and power-law tails \cite{1972PhRvD...5.2439P} are known to dominate at very late times. 
The nonlinear GW memory effect \cite{1992PhRvD..46.4304B, 1991PhRvL..67.1486C, 2010CQGra..27h4036F, 2024CQGra..41v3001M} also accumulates through the merger and ringdown.

Recently, a direct-wave component has also been proposed as a contribution to the merger-ringdown, based on BH-perturbation-theory calculations of a point particle plunging into a larger BH \cite{Oshita:2025qmn} (see also Refs.~\cite{2026PhRvD.113j4013S, 2026arXiv260408680M, 2026arXiv260607173K, 2026arXiv260706677M, 2026arXiv260907326K, 2026arXiv260823209S, 2026arXiv260708570H, 2026Natur.655...71L}).
The merging BHs observed by LIGO and Virgo are of comparable size, so conclusions drawn from perturbation theory necessarily rely on an effective-one-body description of the merger.
Mathematically, the direct wave is the part of the nonlinear merger and subsequent perturbative ringdown waveform (i.e.\ while and after the particle crosses the light ring) with a one-to-one relationship between the time experienced by the plunging particle and the time experienced by the observer.
It models the part of the signal that propagates along null geodesics from the plunging source to an observer without back scattering off the curved background spacetime (hence, \textit{direct}).
Because it tracks the plunging particle, the frequency of the direct wave, it is argued, can provide a probe of geodesics inside the light ring.
In fact, as the particle approaches the event horizon, its frequency asymptotes to the horizon-mode frequency ($m\Omega_H - i\kappa$; see Sec.~\ref{sec:DWfreq}).
Based on this, it has been suggested that identifying a direct wave in real GW signals could provide a way of probing physics inside the light ring, perhaps even down to the event horizon \cite{2026Natur.655...71L}.
A model based on these ideas has also been used to try to measure the increasing horizon area \cite{2026arXiv260606592K}.
If this model were accurate, it would contradict the standard picture described above.

This paper follows Refs.~\cite{Kankani:DWBOB, Kankani:2026kst, 2026arXiv260625021D, 2026arXiv260829466H} in arguing against the picture described in the previous paragraph.
We describe the results of three independent strands of inquiry that suggest a different interpretation of the non-QNM, early-time part of the ringdown signal.
\begin{itemize}
    \item Sec.~\ref{sec:DWfreq} describes the results of measuring the ringdown frequency of NR waveforms from which the QNMs have been removed. The remaining early part of the merger-ringdown is well described by a single damped sinusoid with a non-QNM frequency that is related to the light ring.
    \item Sec.~\ref{sec:BOB} shows that this prompt, non-QNM part of the merger-ringdown is accurately captured by the Backwards One-Body (BOB) model, an analytical model based entirely on light-ring physics; further, this non-QNM part exists across an extended period of time in the merger and early ringdown. This provides further evidence that what has been called the direct wave, hereafter and more properly called the non-QNM part of the merger-ringdown, does not originate from significantly inside the light ring. 
    \item Sec.~\ref{sec:horizonlessUCOs} presents the results of analyzing perturbative ringdown signals on exotic, ultracompact objects with light rings but without event horizons.
    These ringdown signals exhibit similar-looking prompt, non-QNM parts and provide further evidence that this is closely associated with the existence of a light ring.
    This helps to motivate ongoing efforts to produce fully nonlinear simulations of merging ultracompact, horizonless objects.
\end{itemize}

Although our main result is, in a sense, negative---there is no evidence for an observable direct wave originating from inside the light ring---our results also point in a productive direction.
Taken together, recent results in the literature, including those presented here, provide strong evidence for a loud, measurable component of the merger-ringdown that is not due to QNMs.
Like the QNMs, this component is simple to model, with a predictable and physically interpretable frequency associated with the remnant light ring.
Including this contribution, for example by employing BOB, could improve waveform models for future ringdown analyses by allowing them to go beyond QNMs, thereby increasing our ability to test fundamental physics with GWs.

\section{The Frequency of the Non-QNM Part of the Ringdown} \label{sec:DWfreq}

Early analyses of GWs in point-particle models identified the prompt non-QNM component as a signal modulated by the orbital dynamics and only partially screened by the gravitational potential of the remnant \cite{Oshita:2025qmn}, hence transmitting information from inside the light ring out to infinity.  

In Ref.~\cite{2026Natur.655...71L} this component was identified in GW250114 and associated with the properties of the event horizon, namely the rotation frequency $\Omega_H$ and surface gravity $\kappa$. 
In Ref.~\cite{2026arXiv260823209S} the authors add that its complex frequency evolution approaches the horizon properties only at late times, with full evolution also capturing plunge dynamics. 

\begin{figure*}[t]
\begin{minipage}[b]{0.48\linewidth}
    \centering
    \includegraphics[width=\textwidth]{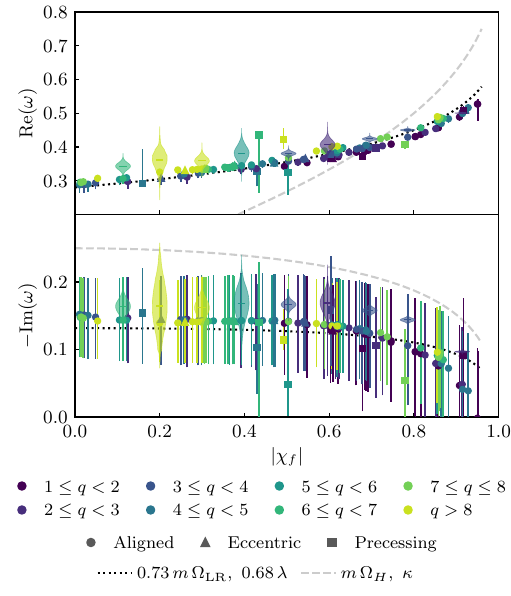}
\end{minipage}
\hspace{0.5cm}
\begin{minipage}[b]{0.48\linewidth}
    \centering
    \includegraphics[width=\textwidth]{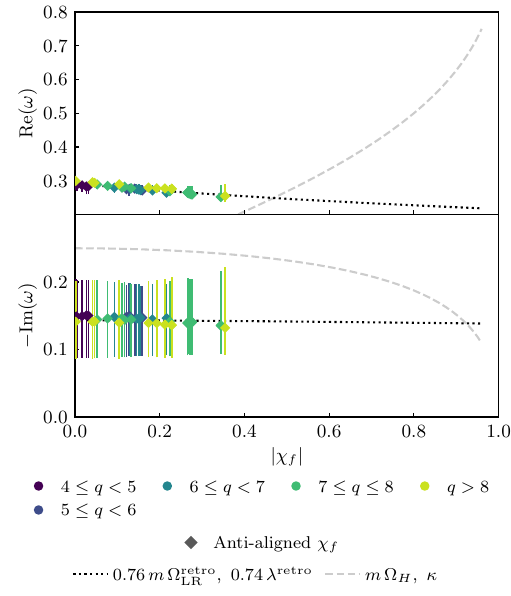}
\end{minipage}
    \caption{ \label{fig:spin_dependencies}
        The spin dependence of the real (top) and imaginary (bottom) parts of the non-QNM frequency are shown for 125 (left) and 43 (right) binary black hole simulations across a range of mass ratios in the $(2,2)$ spherical mode. 
        The figure on the left contains results from simulations with a wide range of mass ratios, spins, and eccentricities; it includes both precessing and non-precessing configurations. These mergers primarily excite prograde modes of the remnant BHs. The figure on the right contains the spin-aligned simulations that have a negative remnant spin (i.e. anti-aligned with the inspiral orbital angular momentum) and which therefore primarily excite retrograde modes of the remnant.
        Behind the lines are two characteristic remnant frequencies: the horizon mode and the prograde (retrograde) light ring, with an overall scaling to fit the line in the latter case. 
        Error bars show the maximum and minimum values of the non-QNM instantaneous frequency of the news across the interval $t \in [-7.5, 7.5]M$ and the marker is the value at $t=0$. 
        The colored violins show the posteriors on the real and imaginary parts of the frequency obtained from a Bayesian free frequency fit as described in Ref.~\cite{Dyer_in_prep}. 
}
\end{figure*}

Ref.~\cite{Kankani:2026kst} identified a coincidence, also pointed out in Ref.~\cite{2026arXiv260823209S}, whereby the frequency evolution of the non-QNM component from a $\chi \sim 0.7$ remnant appears to track the horizon frequency $\Omega_H$ even at early times, far from the regime where frame dragging would dominate. 

Regardless of its origin, these earlier works highlight a clear dependence of the non-QNM component frequency on the remnant spin. 
To disentangle the origins of this component and its relationship with the remnant properties, here we investigate this dependence. 

Earlier works scaled the real and imaginary parts of the non-QNM frequency by the horizon frequency; however this obfuscates a far more obvious relationship, shown in Fig.~\ref{fig:spin_dependencies}, between the non-QNM frequency and the frequency of the light ring. Starting with the upper left panel, Fig.~\ref{fig:spin_dependencies} recreates a version of Fig.~1 of Ref.~\cite{Kankani:2026kst} now using 125 SXS simulations chosen to span a wide range of mass ratios, eccentric as well as quasicircular inspirals, and a range of precessing and non-precessing configurations.
Each waveform was filtered and the instantaneous frequency was determined across the range of times $t \in [-7.5, 7.5] M$ (see App.~\ref{app:sims} for more details on the filtering procedure and simulations used). 
The frequency at $t=0$ (defined to be the peak of the $(2,2)$ strain, even though the frequencies are determined from the news) is shown by a marker and the error bars show the range of frequencies obtained over the stated time range. 
Two lines are also plotted: the first shows the horizon frequency as a function of spin, which bears little resemblance to the measured frequency; the second shows the prograde light ring frequency, scaled by a constant factor $D=0.73$ to best-fit the measured frequencies. 
To obtain this value we fit the frequencies at $t=0$ and include only the aligned-spin simulations in the fit. 

The lower left panel of Fig.~\ref{fig:spin_dependencies} shows the imaginary components of the measured frequencies. In this case we indicate the surface gravity, $\kappa$, and prograde light ring Lyapunov exponent, $\lambda$, with scaling factor $E=0.68$. The imaginary component evolves much more rapidly than the real part. The Lyapunov exponent is the natural counterpart to the light ring frequency, governing the instability timescale for null geodesics escaping the light ring. However, these results do not necessarily indicate a strong relationship between the imaginary part of the non-QNM component and $\lambda$. 

The right panels of Fig.~\ref{fig:spin_dependencies} show the results for 43 non-precessing simulations where the remnant spin points in the opposite direction to the orbital angular momentum in the inspiral. Remnant spins generated in this way cluster at relatively small absolute values, which is why there are no data at spins above $\chi_f \sim 0.4$. 

In the right panels, we fit the non-QNM component frequency to the retrograde light ring frequency, $\Omega_{\rm LR}^{\rm retro}$ and Lyapunov exponent $\lambda_{\rm LR}^{\rm retro}$.  
Both lines are fit separately. 
This choice is motivated by the observation that the retrograde modes of anti-aligned remnant spin black holes are more strongly excited than the prograde modes. 
We observe that the non-QNM frequency also fits this pattern, being strongly correlated to the retrograde light ring. 

There has been much recent discussion on the role of the rational filter in modifying or contributing to the residual non-QNM component. To test whether the rational filter drives the observed behavior, Fig.~\ref{fig:spin_dependencies} also contains nine posteriors from a Bayesian free frequency fit that does \emph{not} use the rational filter and, instead, fixes the QNM frequencies and allows an additional damped sinusoid in the fit to vary. Further details of this method are in App.~\ref{app:sims}. We note that the fitting procedure is highly sensitive to choice of QNMs in the fixed model. This may account for the posterior frequencies being systematically shifted above the best-fit line. 

Fig.~\ref{fig:spin_dependencies} shows the $(2,2)$ mode only; however, a similar relationship holds for $m =3,4$, with a systematic offset for $m=5,6$, as shown in App.~\ref{app:mneq2}. 

\begin{figure}
    \includegraphics[width=\linewidth]{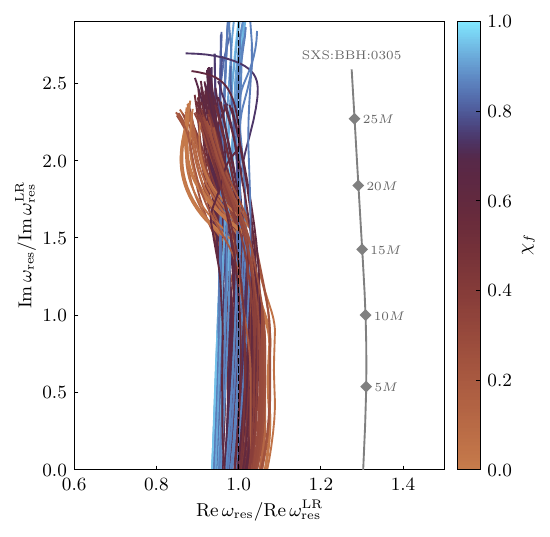}
    \caption{The residual non-QNM component (labeled `res') frequency, scaled by the best-fit light ring model in Fig.~\ref{fig:spin_dependencies} is shown for the 114 SXS simulations with remnant spin aligned to the orbital angular momentum and excluding precessing configurations. The gray line indicates the trajectory of the non-QNM component in SXS:BBH:0305, copied and translated to the right, with corresponding times at which the instantaneous frequency takes that value. Lines end when the amplitude of the non-QNM component reaches $1\%$ of its initial value.}
    \label{fig:dw_evolution}
\end{figure}

In Fig.~\ref{fig:dw_evolution}, we apply the best-fit frequency to a reproduction of Fig.~3 in Ref.~\cite{2026arXiv260823209S}, but using the larger set of simulations from Fig.~\ref{fig:spin_dependencies} (we exclude precessing cases and the 43 `retrograde' simulations from the figure for clarity). 
This choice largely removes the spin dependence and, instead, shows frequencies evolving on practically straight lines for the duration of the non-QNM emission. 

These results support the viewpoint that the non-QNM part of the ringdown is solely a property of the remnant, rather than the progenitor configuration or trajectories. 
Furthermore, rather than the frequency being connected to properties of the horizon, it appears instead to be strongly correlated with the functional dependence of the light ring on spin, up to a rescaling factor of order unity.  

While there is a striking qualitative agreement, we delay an explanation for this relationship until later sections and here only note that this supports the standard picture that properties of the ringdown are associated with excitations around the light ring.

\section{The Analytic Residual for Comparable Mass Systems} \label{sec:BOB}

\input{BOB}

\section{Ringdown without horizons} \label{sec:horizonlessUCOs}

The final strand of evidence we present that the ``direct wave'' component is associated with the light ring of the remnant (and not with its interior or with the horizon) is based on simulations performed on backgrounds corresponding to ultracompact objects (UCOs)---that is, objects that possess light rings but not event horizons.
For this, we use \textit{thin-shell boson stars} (BSs); cf. Refs.~\cite{Boskovic:2021nfs, Marks_2025, Marks:2026yle}. 
Additional details on the construction and properties of these objects can be found in Appendix~\ref{app:thinshell}.
As a toy model for the gravitational ringdown of a UCO, we consider the time evolution of the scalar wave equation on such a background, sourced by a particle falling on a quasi-circular trajectory into the UCO.
We compare this to an evolution performed on a Schwarzschild BH background, as well as on a horizonless background without light rings corresponding to a non-ultracompact BS.
We show that the scalar ringdown for both BH and horizonless UCO spacetimes contains a residual non-QNM component very similar to what is seen in binary BH mergers and has previously been identified with the direct wave. However, the scalar ringdown for the less compact BS spacetimes does not contain this component.

\subsection{Evolving the scalar wave equation} \label{sec:scalar_wave}
First, we describe the numerical scheme used to probe the scalar ``ringdown'' of both BH and UCO spacetimes.
We begin with the static, spherically symmetric metric in polar-areal coordinates,
\begin{equation}
  \dd s^2 = -\alpha(r)^2\,\dd t^2 + X(r)^2\,\dd r^2 + r^2\,\dd\Omega^2 ,
  \label{eq:metric}
\end{equation}
where $\dd\Omega^2$ denotes the usual metric on the two-sphere.
We wish to solve the sourced wave equation for a real scalar field $\psi$, 
\begin{equation}
  \Box\psi \equiv \frac{1}{\sqrt{-g}}\,\partial_\mu\!\left(\sqrt{-g}\,g^{\mu\nu}\partial_\nu\psi\right) = S
  \label{eq:wave}
\end{equation}
with a source function $S$.
We introduce the usual \textit{tortoise coordinate} $\rstar$, defined so that along radial null rays we have $\dd \rstar / \dd t = \pm 1$, i.e. $
    \dd \rstar /\dd r={X}/{\alpha}.
$
We next decompose the source and field into spherical harmonics according to
\begin{equation}
\begin{split}
  \psi&=\sum_{\lm}\frac{\psi_{\lm}(t,r)}{r}\,\Ylm(\theta,\phi),
  \\
  S&=\sum_{\lm}S_{\lm}(t,r)\,\Ylm(\theta,\phi).
  \label{eq:decomp}
\end{split}
\end{equation}
This leads to a modified one-dimensional wave equation for each mode,
\begin{equation}
    \left(-\,\partial_t^2 + \partial_{\rstar}^2 - V_{\ell}(r)\right)\psi_{\lm}
  \;=
  r\,\alpha^2\,S_{\lm}\;
  \label{eq:master}
\end{equation}
where $V_\ell$ is a Regge-Wheeler-like effective potential,
\begin{equation}
  V_{\ell}(r)
  = \frac{\alpha^2}{r^2}\,\ell(\ell+1)
  \;+\; \frac{1}{r}\,\frac{\dd}{\dd \rstar}\!\left(\frac{\alpha}{X}\right).
  \label{eq:Veff}
\end{equation}
We convert Eq.~\eqref{eq:master} to a first-order system by defining $\pi_{\lm} = \dot{\psi}_{\lm} $, then evolve on a grid uniform in $\rstar$ using a fourth-order Runge-Kutta scheme with the method of lines. 
We extract the modes $\pi_{\lm},$ which here provide a scalar analogue of the Bondi news, at a series of specified radii.
At the outer grid boundary we use a simple outgoing Sommerfeld boundary condition, though in practice we take the grid to be large enough that it is out of causal contact with each extraction radius for the duration of our evolutions.

In specifying the source $S$, we would like to approximate a point source that emits scalar radiation while following a trajectory that one might plausibly expect from an extreme-mass-ratio inspiral.
We therefore prescribe a Gaussian profile of peak amplitude $S_0$ and width
$\sigma$ that moves along a trajectory $x_p^\mu(\tau)$ of the background \eqref{eq:metric}, which we take to lie in the equatorial plane so that $\theta_p = \pi / 2$. 
More details are given in Appendix~\ref{app:scalar_evolutions}.

For the trajectory, we integrate a modified version of the geodesic equation, with a four-force $f^\mu$ given by
\begin{equation}
  u^\nu\nabla_\nu u^\mu = f^\mu,
  \qquad
  u_\mu f^\mu = 0 .
  \label{eq:forcedgeo}
\end{equation}
where the latter orthogonality condition ensures that the norm of the four-velocity $u^\mu$ remains constant. 
We choose $f^\mu$ so that its spatial components function as an effective drag whose amplitude is proportional to the local energy density $\rho$, 
\begin{equation}
  f^r=-\gamma\rho u^r,\quad f^\phi=-\gamma\rho u^\phi
  \label{eq:dragansatz}
\end{equation}
where $\gamma > 0$ is an arbitrary constant.
The orthogonality condition then requires
\begin{equation}
  f^t=-\frac{\gamma\rho\bigl[X^2(u^r)^2+r^2(u^\phi)^2\bigr]}{\alpha^2u^t}.
\end{equation}
To obtain quasi-circular trajectories, we begin particles at the areal radius of the ISCO $r = 6M$, with initial angular velocity $\dot\varphi = \Omega - \delta\omega$ where $\delta\omega \ll \Omega$, the ISCO frequency. 

\begin{figure*}
    \includegraphics[width=\linewidth]{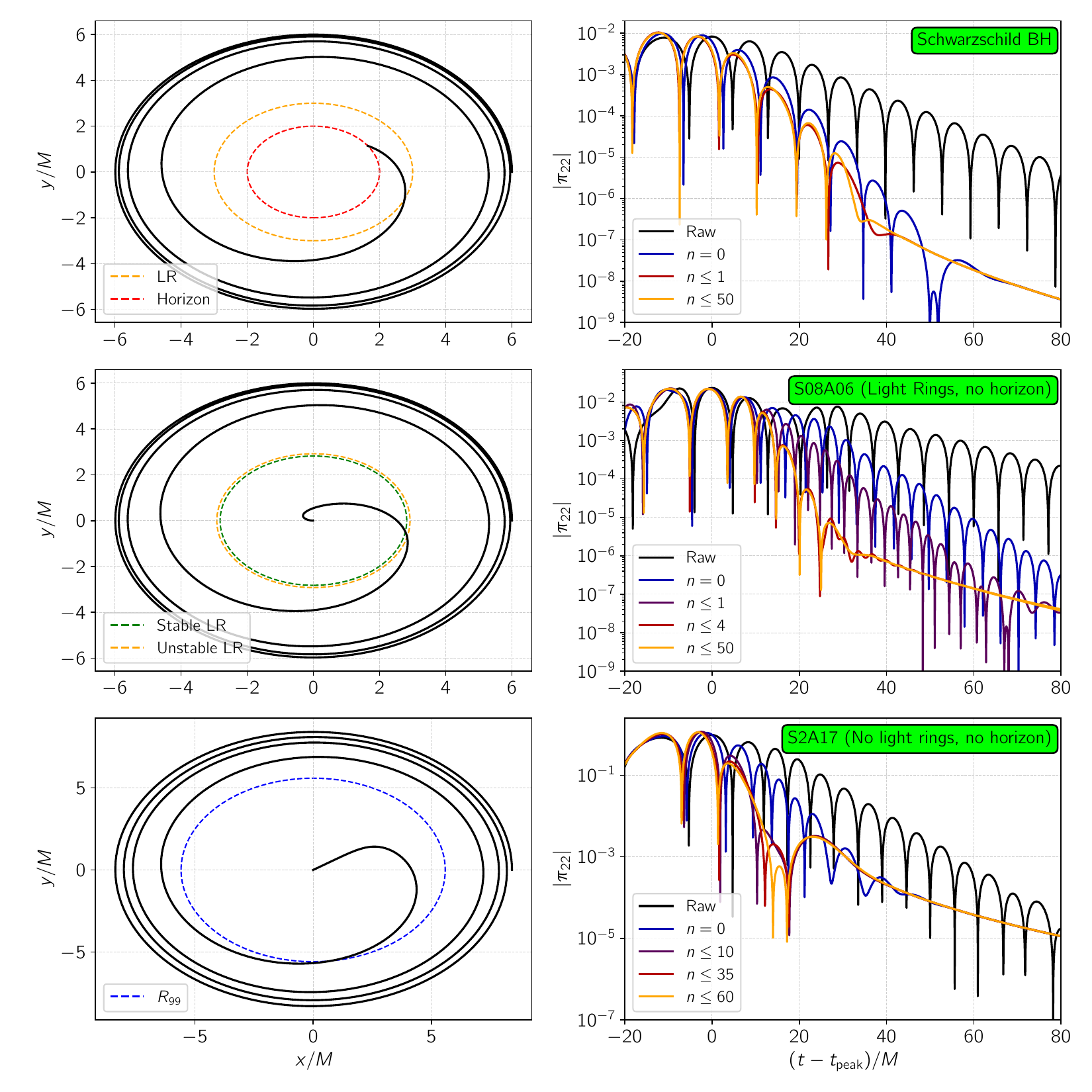}
    \caption{For each main scalar evolution considered in this work, we show the particle trajectory on the left, and both raw and selected filtered waveforms of the $(2,2)$ mode of the real scalar field momentum's $\pi_{22}$---our analogue to the Bondi news--- on the right, normalized by the source's peak amplitude $S_0$. We mark with circles any event horizon or light ring(s); for \texttt{S2A17}, which has neither, we mark the radius $R_{99}$ enclosing $99\%$ of the aspect mass. }
    \label{fig:trajectories_and_waveforms}
\end{figure*}

\subsection{Evolution results}
While our evolutions depend on arbitrary numerical parameters like the drag coefficient $\gamma$ and orbital trajectory, we find that, for reasonable values, the ringdown phase of our scalar waveforms is not sensitive to these choices.
We focus our attention on three representative evolutions: one around a Schwarzschild BH, one around the UCO \texttt{S08A06}, and one around a non-ultracompact BS which we call \texttt{S2A17}.
We give more details on these evolutions in Appendix~\ref{app:scalar_evolutions}.

All three objects considered here are nonspinning, so QNM frequencies are independent of $m$. The filtering procedure is therefore simpler than for the binary black hole simulations: we remove only the $\ell = 2$ modes of each object. For the Schwarzschild black hole these are the scalar QNMs $(2,2,n)$; for the boson stars they are the star's own $\ell = 2$ modes. Since we evolve real scalar fields, each mode must come as a pair, with $\omega_n$ and its mirror $-\omega_n^*$ (the prograde and retrograde modes). The filter removes both so that the filtered signal remains real. For the remainder of this section, ``removing mode $n$'' means removing this pair.
Details on our approach to calculating QNM frequencies for horizonless spacetimes are given in Appendix~\ref{app:computing_qnms}.
For both \texttt{S2A17} and \texttt{S08A06}, we are able to accurately compute $n \ge 50$ overtone frequencies of the $\ell = 2$ modes, which will be our focus.

We show the particle trajectories and extracted waveforms, along with the results of filtration, for each evolution in Fig.~\ref{fig:trajectories_and_waveforms}, along with the results of QNM filtering.
In the BH case, once the $n = 0,1$ modes have been filtered out, there remains a clearly visible non-QNM residual with real frequency not vastly different from that of the first overtone. This residual component does not significantly change under the filtration of higher modes.
We therefore interpret it as a scalar analogue of the component that has heretofore been interpreted as a direct wave.
For the ultracompact \texttt{S08A06} evolution (horizonless but with light rings), a few more overtones must be filtered out, because of their much slower decay rate.
When this is done, however, the conclusion is the same: a non-QNM residual remains, which does not significantly change after filtration of the first $\sim 4$ overtones, and whose real frequency is comparable to that of the first overtone--- although now decaying much more quickly.
Again, this residual can be seen as an analogue to the direct wave.

When we repeat this analysis with an evolution performed using the non-ultracompact BS \texttt{S2A17} (neither an event horizon nor light rings), the residual once again stabilizes after the filtration of $\sim 4$ overtones.
This time, however, it does not appear to be well-described by a simple decaying oscillation, as the residual BH and UCO components were.
Thus, the scalar non-QNM component we have identified appears in our evolutions when--- and only when--- the spacetime has light rings, regardless of whether it contains an event horizon.
We interpret this as additional evidence that the ``direct wave'' component is ultimately related to the light ring, and not the event horizon.

\section{Discussion and Conclusion} \label{sec:discussion}
The ringdown following the merger of two approximately equal-mass BHs is dominated by QNMs.
However, QNMs do not make up the whole ringdown signal.
Additionally, there are power-law tails at late times, nonlinear (e.g. quadratic) QNMs at intermediate times, and (as has been shown here but is also, by now, well established in the literature) a prompt, non-QNM component at early times.

This prompt, non-QNM component has a frequency that is related to the light ring of the remnant BH.
This is made clear by the fact that its properties are tightly correlated with the remnant spin.
This was first pointed out by Ref.~\cite{Kankani:2026kst} and has been extensively investigated here in Sec.~\ref{sec:DWfreq}.

The frequency of this non-QNM part of the ringdown cannot be explained by any of the QNMs.
It has a frequency lower than even the lowest, fundamental QNM, and much lower than those of any of the overtones, harmonics, or nonlinear QNMs.
This low frequency may explain why this part of the ringdown was missed by many early studies that focused solely on QNMs.

Motivated by calculations in black-hole perturbation theory, this non-QNM part of the signal has previously been identified with the direct wave.
We find no evidence to support this claim.
On the contrary, while a direct-wave interpretation would imply a signal produced from inside the light ring by a particle plunging towards the horizon, we have shown several independent strands of evidence that suggest it is instead sourced at the light ring.
We have shown that the frequency of the non-QNM part of the ringdown is proportional to the frequency of the light ring (Sec.~\ref{sec:DWfreq}), that the properties of the non-QNM part can be understood with reference to an analytical merger-ringdown model explicitly based only on light-ring physics (Sec.~\ref{sec:BOB}), and that morphologically similar non-QNM ringdown signals are produced by ultracompact objects with light rings but without event horizons (Sec.~\ref{sec:horizonlessUCOs}).


\phantom{.}
\section*{Author Contributions}

The author list is alphabetical.
The authors all contributed equally to this work and writing the manuscript.
RD was primarily responsible for the filtering analysis in Sec.~\ref{sec:DWfreq}.
AK was primarily responsible for the analytic calculations analysis in Sec.~\ref{sec:BOB}.
GM was primarily responsible for the boson star calculations in Sec.~\ref{sec:horizonlessUCOs}.

\section*{Data and Code Availability}

The data and code to reproduce the results in Sec.~\ref{sec:DWfreq} are available at this repository \href{https://github.com/Richardvnd/direct_waves_LR}{\faGithub}.
The data and code to reproduce the results in Sec.~\ref{sec:BOB} is available at this repository \href{https://github.com/AnujKankani/BOB_analytical_rational_filter}{\faGithub}
The data and code to reproduce the results in Sec.~\ref{sec:horizonlessUCOs} are available at this repository \href{https://github.com/GarethAMarks/UltraWave}{\faGithub}.

\begin{acknowledgments}
    We are grateful to Adrian Ka-Wai Chung, Suchindram Dasgupta, and Ulrich Sperhake for useful discussions about this work.
    RD acknowledges support from the Science and Technology Facilities Council (Grant No. ST/Y509139/1).
    GAM is supported by the Cambridge Trust at the University of Cambridge.
    AK and STM were supported in part by NASA grants 22-LPS22-0022 and 24-2024EPSCoR-0010.
    The calculations in this paper made use of the following software packages: \texttt{NumPy} \cite{harris2020array}, \texttt{SciPy} \cite{2020SciPy-NMeth}, \texttt{Matplotlib} \cite{Hunter:2007}, \texttt{SphericalBSEvolver} \cite{Marks_Spherical_BS_Evolver}, \texttt{SymPy}\cite{sympy}, \texttt{mpmath} \cite{mpmath}, \texttt{SXS} \cite{Boyle2025SXS}, \texttt{qnm} \cite{Stein:2019mop}, \texttt{kuibit} \cite{kuibit}, \texttt{gwBOB} \cite{gwBOB}. Computations were done on the Swirles cluster at the University of Cambridge. Computational resources were provided by the WVU Research Computing Harpers Ferry HPC cluster, which is funded in part through a NASA Congressionally Directed Spending.
    The large language models (LLMs) \texttt{Claude Opus 5.0} and \texttt{Claude Opus 5.5} were used in preparing this work. Both models assisted with the BOB derivation, particularly by identifying many helpful identities; every step of the derivation was manually verified by A.K. Both models were also used to write some parts of the analysis and plotting scripts for Secs.~II, ~III and~IV, under close supervision by R.D., A.K. and G.A.M.

\end{acknowledgments}


\bibliographystyle{apsrev4-2}
\bibliography{references}


\appendix

\section{SXS simulations and QNM choice} \label{app:sims}

This work uses NR waveforms from the Spectral Einstein Code (SpEC), found in the Simulating Extreme Spacetimes catalog \cite{SXS_CCE_catalog, boyle_sxs_2019}. In order to span a wide range of parameter space, we primarily use the large number of available extrapolated waveforms. In Sec.~\ref{sec:DWfreq} we also include the 13 publicly available simulations with waveforms obtained using Cauchy-characteristic evolution (CCE). These waveform data were created using the SpECTRE CCE module \cite{Moxon2023, spectrecode, PhysRevD.102.044052}. Currently these simulations only sparsely cover a range of progenitor parameters, however we include them to ensure our results are consistent with these, significantly more accurate, waveforms. They are post-processed as described in Ref.~\cite{Dyer:2025iwj}. 

Fig. \ref{fig:spin_dependencies} uses a total of 125 SXS simulations (both extrapolated and CCE) on the left axes and an additional 43 simulations with a remnant that is anti-aligned to the inspiral orbital angular momentum on the right. 
The 125 simulations are chosen to include all used in Ref. \cite{2026arXiv260823209S}, thereby also including all the simulations from \cite{Kankani:2026kst} except SXS:BBH:2095, which was replaced by SXS:BBH:3584 in Ref.~\cite{2026arXiv260823209S}. For completeness we also include the three precessing simulations from the CCE catalogue that were omitted in \cite{2026arXiv260823209S}. The remaining 96 simulations are chosen so that there are 10 further aligned simulations in each mass range specified on the figure, in addition to one precessing and eccentric configuration in each. 
The 43 anti-aligned simulations were obtained from an exhaustive search of the SXS catalog. 
A full list of simulations is provided in the linked Git repository \href{https://github.com/Richardvnd/direct_waves_LR/blob/main/simulations.txt}{\faGithub}. 

In Figs.~\ref{fig:spin_dependencies} and \ref{fig:dw_evolution} SXS simulations were filtered according to the prescription given in Ref.~\cite{2026arXiv260823209S}. The $n = 0, 1, 2$ prograde and retrograde QNMs, for every $2 \leq \ell \leq 6$ were filtered from the Bondi news in each waveform. 

To obtain the Bayesian free frequency fits a set of fixed-frequency QNMs were included in the fit, in addition to a damped sinusoid whose frequency was allowed to vary with flat priors on both the real and imaginary parts in the intervals $(0.25, 2.5)$ and $(0.05, 20.0)$ multiplied by the best-fit, scaled light ring frequency values obtained in Fig.~\ref{fig:spin_dependencies}. 
For every fit, the QNMs $(2,2,n,\pm)$ for $n \leq 1$ and $(3,2,0,+)$ were used.  
These models are non-linear and have non-Gaussian posteriors that are sampled using more expensive Markov chain Monte Carlo methods (MCMC); therefore they were only performed for nine simulations at a start time of $t_0 = 0M$ with a fitting window of $T = 15M$. The simulations were selected to span the range of remnant spins and are flagged in the full list of simulations in the Git repository. The numerical uncertainty on each simulation was estimated by computing the residual between the highest two levels available. We used the maximum absolute residual over $0 \leq t \leq 130M$, after aligning the two levels in time and phase before merger, as the standard deviation of a conservative white-noise model. MCMC chains were run for 200,000 steps with 48 walkers. 30\% of the chain was removed as burn-in. Full details of the Bayesian free frequency fitting approach are given in Ref.~\cite{Dyer_in_prep}.

\section{The non-QNM component frequency in $m \geq 3$ modes} \label{app:mneq2}

\begin{figure*}
    \includegraphics[width=\linewidth]{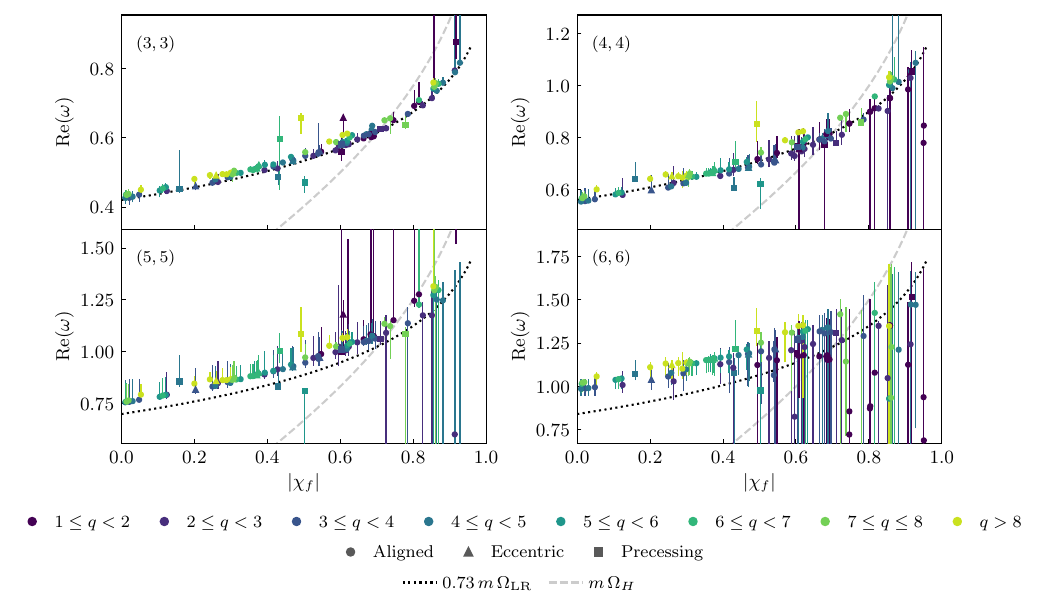}
    \caption{The spin dependence of the real parts of the non-QNM frequency in $3 \leq m \leq 6$ are shown for the binary black hole simulations as in Fig.~\ref{fig:spin_dependencies}. The $(3,3)$ and $(5,5)$ panels do not include the 12 equal-mass, equal-spin simulations, in which odd-m modes are suppressed by symmetry. The non-QNM instantaneous frequencies are plotted, with the marker indicating the frequency at $0M$ and error bars indicating the maximum and minimum frequencies obtained in the interval $t \in [-7.5, 7.5]M$. The dashed line shows the horizon frequency and the dotted line shows the best-fit light ring frequency obtained from fitting the (2,2) mode in Fig.~\ref{fig:spin_dependencies}.}
    \label{fig:spin_dependence_highm}
\end{figure*}

Here we investigate the non-QNM component for $3 \leq m \leq 6$. In Fig.~\ref{fig:spin_dependence_highm} we plot the real part of the frequency as in Fig.~\ref{fig:spin_dependencies} and compare it to the best-fit line obtained there. For modes $m = 3$ and $4$ the fit holds relatively well. In $m = 5$ and $6$ the curve deviates systematically from the line. This may be indicative of either other factors that are not accounted for in this simplified phenomenological fit, or an overall loss of accuracy in constraining the non-QNM component frequency in these quieter modes. 

\input{simplified_BOB}

\input{BOB_appendix}

\input{BOB2_appendix}

\section{Thin-shell boson stars} \label{app:thinshell}
In this appendix, we give more details on the underlying matter model and geometric properties of our UCO spacetimes.
To obtain a simple example of a spherically symmetric spacetime with light rings but without event horizons, we use the \textit{thin-shell}  boson stars (BSs) as described in Refs.~\cite{ Boskovic:2021nfs, Collodel:2022jly}.
These are equilibrium solutions derived from a minimally coupled massive complex scalar field $\varphi$ with action given by
\begin{equation}
    \!\!\!
    \mathcal{S} = \int \du ^4x\frac{\sqrt{-g}}{2}\left\{\frac{R}{8 \pi}  -\left[
  g^{\mu\nu}\nabla_{\mu}\bar{\varphi}\,\nabla_{\nu}\varphi
  +V(|\varphi|^2)\right] \right\}\, ,\label{eq:action}
\end{equation}
where we use the so-called \textit{solitonic} potential, \begin{equation} \label{eq:solitonic}
    V(|\varphi|^2)= \mu^2 |\varphi|^2 \left(
  1-2\frac{|\varphi|^2}{\sigma_0^2}
  \right)^2,
\end{equation}
where $\mu$ is the scalar mass and $\sigma_0 > 0$ a free parameter.
The presence of a degenerate vacuum state where $|\varphi| = \sigma_0 / \sqrt{2}$ allows the field to assume highly compact equilibrium configurations while remaining perturbatively stable.
Indeed, the solitonic BS we will focus on here was previously used to provide the first long-term dynamically stable evolutions of a horizonless object with light rings in full numerical relativity~\cite{Marks_2025}, making this particularly well-motivated as a black hole mimicker.
Details on the construction of these models can be found in Ref.~\cite{Marks_2025_B}.

Here, however, we will only be interested in the resulting background geometry~\eqref{eq:metric}. Any \textit{ultracompact object} (UCO)--- which we use here to mean an object possessing a light ring, but not an event horizon--- must in fact possess \textit{two} light rings, given a broad set of assumptions \cite{cunha_light-ring_2017, Cunha:2020azh}.

The outer light ring is unstable, corresponding roughly (in the spherically symmetric case) to the light ring at areal radius $r = 3M$ for a Schwarzschild BH. 
The additional light ring is stable, meaning that null geodesics originating close to the orbit may remain indefinitely bound to it.
We will focus on two UCOs, labeled \texttt{S08A06} and \texttt{S06A044}, respectively, in  Ref.~\cite{Marks_2025}. \texttt{S08A06} has its stable light ring located at areal radius $r = 2.82M$, and its unstable light ring at $r = 2.92M$, while \texttt{S06A044} has its stable light ring located at areal radius $r = 2.70M$, and its unstable light ring at $r = 3.00M$.

\begin{figure}
    \includegraphics[width=\linewidth]{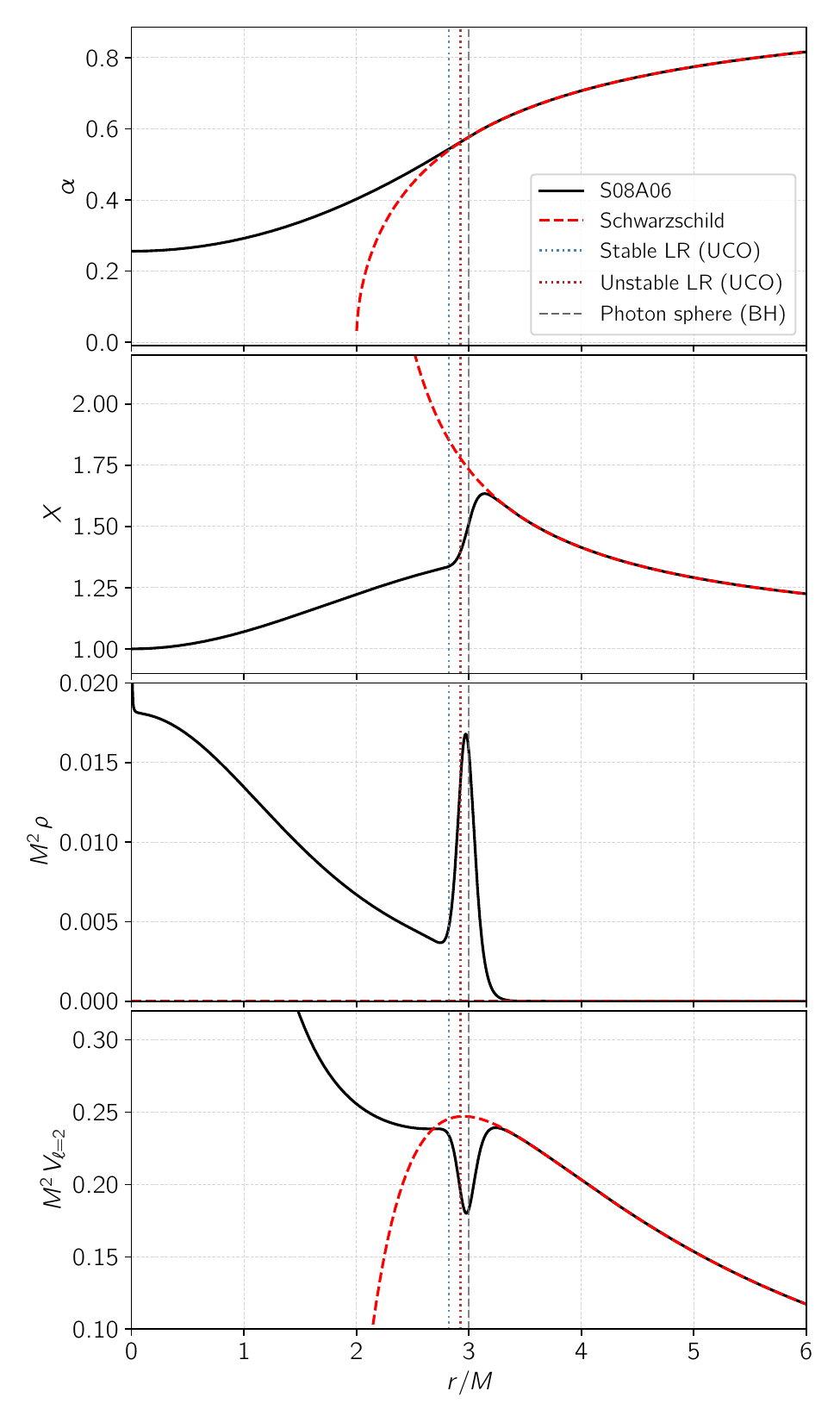}
    \caption{Profiles of geometric quantities associated with the UCO \texttt{S08A06}, compared against a Schwarzschild BH. From top to bottom: lapse $\alpha$, $(r,r)$ metric component $X$, energy density $\rho$, and effective potential governing the propagation of $\ell = 2$ scalar waves $V_{\ell = 2} $ (see Eq.~\eqref{eq:Veff}). We mark the areal radii of the stable and unstable light rings associated with the UCO and BH by vertical lines.}
    \label{fig:UCO_geometry}
\end{figure}

Unlike (for example) Tolman-Oppenheimer-Volkoff stars and gravastars, BSs do not have a hard surface. 
Instead, the scalar amplitude $|\varphi|$ decays exponentially in the asymptotic region.
The decay is particularly rapid for the thin-shell models we here consider, which take their name from the fact that their energy density profiles are dominated by a steep gradient term at finite radius.
In Fig.~\ref{fig:UCO_geometry}, we compare the profile of \texttt{S08A06}'s geometry to that of a Schwarzschild BH.

\section{Additional details on the scalar evolutions} \label{app:scalar_evolutions}
In this appendix, we give more details on the numerical framework and parameter choices used for our scalar wave evolutions.
First, we discuss in more detail our choice of source function $S$.
We will be interested in the limit $\sigma \ll M$, so for convenience define the Gaussian profile with respect to coordinate distance,
\begin{equation}
  S=S_0\,\exp\!\left(-\frac{r^2+r_p^2-2\,r\,r_p\cos\phi_p}{2\sigma^2}\right),
  \label{eq:gauss}
\end{equation}
where, as before, $x_p^\mu(\tau)$ denotes the coordinates of the particle trajectory. Modes of the source can then be computed analytically, according to
\begin{equation}
  S_{\lm}=4\pi S_0\,
  e^{-[\,r^2+r_p^2\,]/2\sigma^2}\,
  i_\ell\!\left(\frac{r\,r_p}{\sigma^2}\right)
  Y_{\ell m}^{\ast}\!\bigl(\tfrac{\pi}{2},\phi_p\bigr),
  \label{eq:Slm}
\end{equation}
with $i_\ell$ the modified spherical Bessel function of the first kind. 
To account for geometric distortion, the width $\sigma$ is dynamically updated on each timestep to ensure conservation of the ``charge,''
\begin{equation}
    Q:= u_p^t \int \dd^3 x \sqrt{-g}S.
\end{equation}
For horizonless spacetimes, we also include the option to dynamically evolve $\sigma$ according to
\begin{equation}
  \frac{\dd\sigma}{\dd\tau} = \kappa_0\left(\frac{\rho(r_p)}{\rho_{\rm ref}}\right)^{p}
  \bigl(\lambda_{\rm mean} -\sigma\bigr),
  \label{eq:sigma_relax}
\end{equation}
where $\lambda_{\rm ref}$ is a reference scale taken to be on the order of the spacetime mass, $\rho$ is the energy density with mean value $\rho_{\rm mean}$ over the stellar profile, and $p$ and $\kappa_0$ are free parameters controlling the rate at which the source's length scale relaxes to that of the background after merger.
A comprehensive list of the parameter choices made in our three primary evolutions is given in Table~\ref{tab:param_choices}, though we emphasize that the qualitative features shown in Fig.~\ref{fig:trajectories_and_waveforms} do not depend significantly on these choices.
We have also checked that our numerical code converges at the expected fourth order with increasing resolution.
\begin{table}[t]
\centering
\caption{Parameter choices in our three primary evolutions (cf. Fig.~\ref{fig:trajectories_and_waveforms}): the background, Gaussian width $\sigma$, offset of the original angular frequency $\delta\omega$ from that corresponding to a stable circular orbit $\Omega$, drag coefficient $\gamma$ and relaxation parameters $\kappa_0$ and $p$. Note that the latter three are irrelevant for a vacuum spacetime.}
\label{tab:param_choices}
\begin{tabular}{@{}cccccc@{}}
\toprule
\textbf{Background} &$\sigma / M$ &$\delta\omega/\Omega $ &$\gamma$   &$\kappa_0$  & $p$     \\ \midrule
 Schwarzschild &0.02  &0.01  &--  &--  &--  \\
 \texttt{S08A06} &0.02  &0.01  &50  &0.01  &0.1   \\ 
 \texttt{S2A17} &0.04 &0.002 &25 &0.01  &0.1    \\ 
\bottomrule
\end{tabular}
\end{table}

\section{Computing the quasinormal modes} \label{app:computing_qnms}
In this Appendix, we discuss the calculation of the QNM spectrum for our horizonless, and in particular ultracompact, spacetimes.
To determine all the components present in our waveforms, we wish to filter out all scalar QNM frequencies present.
The QNM spectra of UCOs, however, can differ markedly from those of ordinary BHs~\cite{Franzin:2023slm}. 
In spherical symmetry, comparisons of the ringdown have found that, at early times, UCO signatures can closely imitate those of a Schwarzschild BH with clear deviations--- such as the so-called \textit{echoes}--- only becoming apparent late in the signal ~\cite{Cardoso:2016rao, Cardoso:2016oxy}.
Qualitatively, this can be understood in terms of the effective potential governing the dynamics of null geodesics.
The presence of the stable light ring for UCOs corresponds to a local minimum in this potential, creating an effective cavity in the object's interior which can lead to QNMs with very slow decay rates relative to their BH analogues.
When the cavity is deep, these slow-decaying modes tend to be excited much less than modes closer to the Schwarzschild spectrum, helping to explain the appearance of echoes in these signatures \cite{Rosato:2025lxb}.
Our choice of \texttt{S08A06} as our primary UCO is motivated by the relative \textit{shallowness} of its cavity, which reduces the number of overtones we must filter before the residual stabilizes, and raises our confidence that this residual cannot be explained by a QNM frequency missing from our computed values.

\begin{figure}[t!]
    \includegraphics[width=\linewidth]{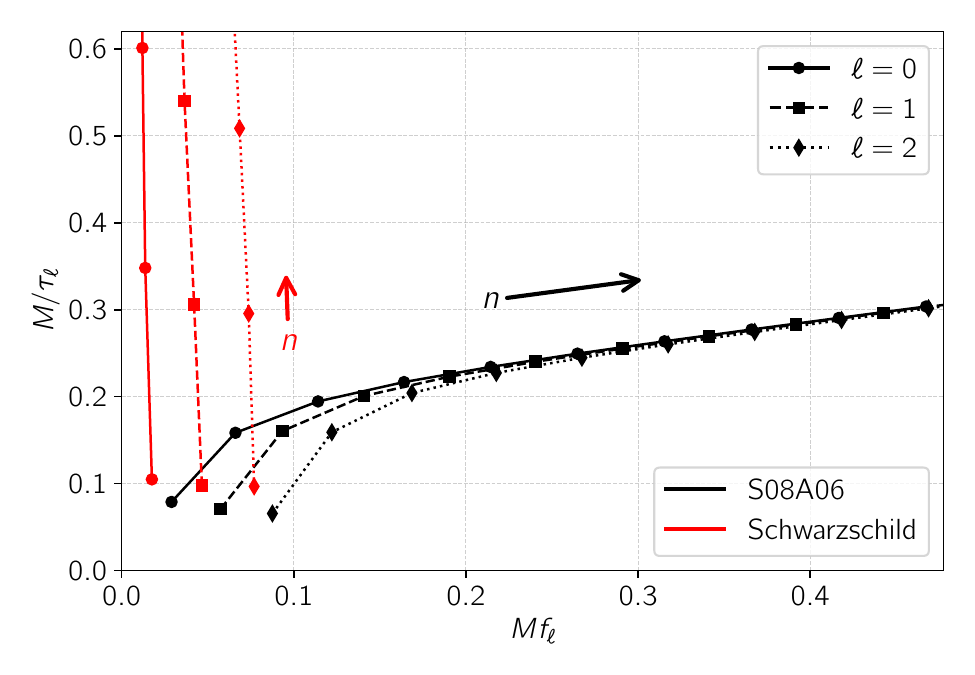}
    \caption{Representation of the spin-zero $\ell = 0,1,2$ QNM spectrum for the UCO \texttt{S08A06} compared to that of a Schwarzschild BH. We give each complex QNM frequency $\omega_\ell$ in terms of the real frequency $f_\ell$ and damping time $\tau_\ell$ so that $\omega_\ell = 2\pi f_\ell - \iu / \tau_\ell$. The direction of increasing overtone number $n$ is indicated with an arrow for each object.   }
    \label{fig:UCO_QNMs}
\end{figure}

To compute the (spin-$0$) QNM spectrum of our horizonless spacetimes, we adapt the methods described in Ref.~\cite{Macedo:2016wgh}.
In particular, we adopt a two-domain approach, using direct integration in the BS interior and a continued fraction method in the exterior, where deviations from the Schwarzschild spacetime are negligible.
We must therefore continuously match these two solutions at some radius to ensure all boundary conditions are simultaneously satisfied.
In practice, we find that any matching radius roughly between the outermost light ring at $r \sim 3M$ and the ISCO at $r \sim 6M$ allows us to compute QNMs with uncertainties of $\sim 0.001\%$ for the ground mode, rising to $\sim 0.01\%$ for the first overtones and gradually to $\sim 1\%$ for the highest overtones we compute (typically around $n = 50$).

In Fig.~\ref{fig:UCO_QNMs}, we compare the QNM spectrum for the UCO \texttt{S08A06} to that of a Schwarzschild BH. 
Notice that while the fundamental modes are comparable, the overtones rapidly depart, with the UCO overtones assuming much higher frequencies and slower decay rates than their Schwarzschild counterparts.

\end{document}

%% file: BOB.tex
Most studies involving the rational filter---a rational function of frequency designed to remove specific complex frequencies---for comparable mass systems have applied it numerically to NR waveforms \cite{ma2022quasinormal,Oshita:2025qmn,2026Natur.655...71L,Kankani:DWBOB,Kankani:2026kst,2026arXiv260823209S}. Analytical treatments have been limited to the point particle case \cite{Oshita:2025qmn,2026arXiv260706677M} or toy models \cite{ma2022quasinormal} and how far those insights transfer to the comparable mass case remains unclear. Here, we close this gap by applying the rational filter analytically to the Backwards One Body (BOB) model \cite{mcwilliams2019analytical,kankani2026analyzing,Kankani:DWBOB}, a light ring based model for the merger-ringdown radiation. This provides a theoretical understanding of the ``loud'' early time portion of the non-QNM component. Our analysis focuses on the filter's behavior before and near the peak of the unfiltered news, the first time derivative of the strain. We present an exact analytical expression for the BOB non-QNM component that accurately models the NR non-QNM component for an extended period of time, as far as $\sim20M$ after the peak for the highest remnant spin cases. Furthermore, building on methods introduced in \cite{2026arXiv260829466H}, we decompose the BOB non-QNM component into the unfiltered BOB waveform and the ``filter effects'', providing insight into when the non-QNM component is ``revealing'' a subdominant component, and when it is largely redistributed information from later times.

\subsection{The Backwards One Body Model}
Derived from the perturbation of null geodesics about the light ring of the remnant black hole \cite{ferrari,mcwilliams2019analytical}, or equivalently through the approximate summation of QNM information \cite{Kankani:DWBOB}, BOB provides a physically motivated, highly accurate, and minimally tuned model for the merger-ringdown radiation \cite{kankani2026analyzing}. These properties make it a useful model for understanding the effects of the rational filter for comparable mass systems. It retains a high level of accuracy all the way back to the peak of the waveform, while still allowing us to understand the exact pole structure in the complex plane \cite{kankani2026analyzing,Kankani:DWBOB}. 

BOB fundamentally models the gravitational wave news \cite{Kankani:DWBOB} and predicts a hyperbolic secant profile for the amplitude, governed by the damping time of the fundamental quasinormal mode, $\tau = -1/\Im(\omega_{220})$:
\begin{align}
    A_\bob(t) = A_p\, \text{sech}\bigg(\frac{t-t_p}{\tau}\bigg),
    \label{eq:BOB_amplitude}
\end{align}
The frequency is derived through the relationship between the amplitude and frequency developed in \cite{baker2008mergers,mcwilliams2019analytical}. Unfortunately, the complexity of the resulting phase evolution makes us unable to find a closed form expression for the Fourier transform of BOB. Therefore, we propose a slight modification to the frequency evolution to make the following calculations analytically tractable. Details of this approximation and comparison to NR waveforms are provided in Appendix~\ref{app:BOB_simple_freq}. 

The simplified frequency model is given by 
\begin{equation}
    \Omega_\bob = \frac{\Omega_{\qnm} + \Omega_0}{2} + \frac{\Omega_\qnm-\Omega_0}{2}\tanh{\bigg(\frac{t-t_p}{\tau}\bigg)}.
     \label{eq:bob_freq_simple}
\end{equation}
The waveform frequency $\omega$ is related to $\Omega$ by $\omega = m\Omega$, where $m$ is the gravitational wave mode number. $t_p$ is the time of the peak amplitude of the waveform. $\Omega_\qnm = \Re(\omega_{\qnm})/m$ gives the asymptotic limit of the frequency, where $\omega_{\qnm}$ is the true Kerr complex QNM. $\Omega_0$ is the initial frequency parameter, defined at $t=-\infty$, well before BOB becomes accurate. Appendix~\ref{app:BOB_simple_freq} provides further details on our simplification and $\Omega_0$.

With our simplified frequency model we can now obtain a frequency domain BOB waveform. First we define the variable $s = (t-t_p)/\tau$. Integrating Eq.~\eqref{eq:bob_freq_simple} in the time domain gives us the simplified phase evolution. Defining $\lambda = (m\tau/2)(\Omega_\qnm - \Omega_0)$ and $\bar{\Omega} = (\Omega_\qnm+\Omega_0)/2$ we can rewrite the ``simplified'' time domain BOB waveform as
\begin{equation}
    \mathcal{N}_\bob (t) = A_p\text{sech}(s)^{1+i\lambda}e^{-im\bar{\Omega}(s\tau+t_p)}.
    \label{eq:BOB_simple_time_domain}
\end{equation}
The Fourier transform of Eq.~\eqref{eq:BOB_simple_time_domain} gives the frequency domain BOB model:
\begin{equation} \label{eq:BOB_fd}
        \mathcal{N}_{\rm BOB}{\left(\omega \right)} = \frac{A_{p} \tau 2^{i \lambda}  e^{i\nu t_p/\tau} \Gamma(\frac{1 + i \lambda - i\nu}{2}) \Gamma (\frac{1 + i \lambda + i \nu}{2})}{\Gamma (1 + i \lambda)},
\end{equation}
where $ \nu = (\omega-m\bar{\Omega})\tau$.

\subsection{The Rational Filter}

The full Kerr prograde rational filter is given by
\begin{equation}
    F_p(\omega) = \prod_{n=0}^{N} \frac{\omega - \omega_{lmn}}{\omega - \omega^*_{lmn}}
    \label{eq:kerr_rational_filter}
\end{equation}
where $\omega_{lmn}$ are the complex Kerr QNM frequencies and $\omega$ is our waveform frequency.
However, by writing out the hyperbolic secant amplitude profile, Eq.~\eqref{eq:BOB_amplitude}, as a series
 \begin{align}
    A_p\text{sech}\left(\frac{t-t_p}{\tau}\right) = 2 A_p \sum_{n=0}^{\infty} (-1)^n e^{-(2n+1)\frac{t-t_p}{\tau}} \notag \\
    = 2A_p\bigg(e^{-\frac{t-t_p}{\tau}} - e^{-3\frac{t-t_p}{\tau}} + e^{-5\frac{t-t_p}{\tau}} + \cdots\bigg),
    \label{eq:amplitude_expansion}
\end{align}
it becomes clear that BOB incorporates the overtone damping times through the eikonal approximation where
\begin{equation}
    \tau_n = \frac{\tau_{n=0}}{2n+1}.
\end{equation}
Therefore, the Kerr rational filter in Eq.~\eqref{eq:kerr_rational_filter} will not exactly cancel the poles present in the hyperbolic secant. To cancel out the poles in BOB exactly, we use an ``eikonal rational filter'' where the overtones ($n>0$) are given in the eikonal limit.
\begin{equation}
    F_{p,\bob}(\omega) = \prod_{n=0}^{N} \frac{\omega - \Re(\omega_{220})+i((2n+1)/\tau_{220})}{\omega - \Re(\omega_{220}) - i((2n+1)/\tau_{220})},
    \label{eq:eikonal_filter}
\end{equation}
where $\tau_{220} = -1/\Im(\omega_{220})$ is the exact Kerr damping time of the fundamental mode. 

\subsection{Applying the Filter to BOB}

\begin{figure}
    \centering
    \includegraphics[width=0.75\linewidth]{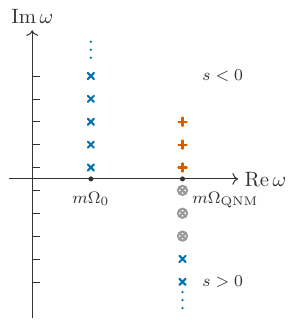}
    \caption{Pole structure after applying the rational filter in Eq.~\eqref{eq:eikonal_filter} to BOB. The unfiltered BOB poles are marked as blue crosses. The eliminated BOB poles are grey circled crosses and the artificial poles introduced by the filter are marked with an orange plus sign. }
    \label{fig:pole_structure}
\end{figure}

With the frequency domain BOB waveform, Eq.~\eqref{eq:BOB_fd}, and our eikonal rational filter, Eq.~\eqref{eq:eikonal_filter}, we now apply the filter in the frequency domain and inverse Fourier transform to obtain the time domain non-QNM component:
\begin{equation}
     \mathcal{N}_{\res,\bob} (t) = \frac{1}{2\pi}\int_{-\infty}^{\infty} F_{p,\bob}(\omega) \mathcal{N}_{\bob} (\omega) e^{-i\omega t} d\omega.
    \label{eq:BOB_filtered_master}
\end{equation}

While the full derivation is given in Appendix~\ref{app:BOB_derivation}, we present the main results here. Because of the length of the equations, several substitutions are made to make the presentation cleaner. The exact substitutions are listed in the beginning of Appendix~\ref{app:BOB_derivation}. We note here that $\kappa_n$ are the $_2F_1$ hypergeometric functions. 

As shown in Fig.~\ref{fig:pole_structure}, Eq.~\eqref{eq:BOB_filtered_master} contains filter and BOB poles in the upper half plane and the uncanceled BOB poles in the lower half plane. Closing the contour in the upper half plane (UHP) gives the solution to Eq.~\eqref{eq:BOB_filtered_master} for $s<0$
\begin{widetext}
\begin{align}
     \mathcal{N}_{\res,\bob, s<0} (t)  = \zeta(s)e^s  \bigg[&z_+\frac{\Gamma(2+N+i\lambda)\Gamma(i\lambda-N)}{\Gamma(1+i\lambda)^2}\kappa_1(s) - z_- (-1)^{N}\frac{\Gamma(i\lambda)\Gamma(2+N)}{\Gamma(1+N)\Gamma(1+i\lambda)}\kappa_2(s)\bigg] 
    \label{eq:BOB_dw_finite_N_sl0}
\end{align}
and closing in the lower half plane (LHP) gives
\begin{align}
\mathcal{N}_{\res,\bob, s>0} (t) = -\zeta(s) z_-(s)(-1)^{N} 
e^{-s(3+2N)} \frac{\Gamma(2+N)\Gamma(2 + i\lambda + N)}{\Gamma(3+2N)\Gamma(1+i\lambda)}\kappa_3(s)
\label{eq:BOB_dw_finite_N_sg0}
\end{align}
\end{widetext}

\begin{figure}
    \centering
    \includegraphics[width=\linewidth]{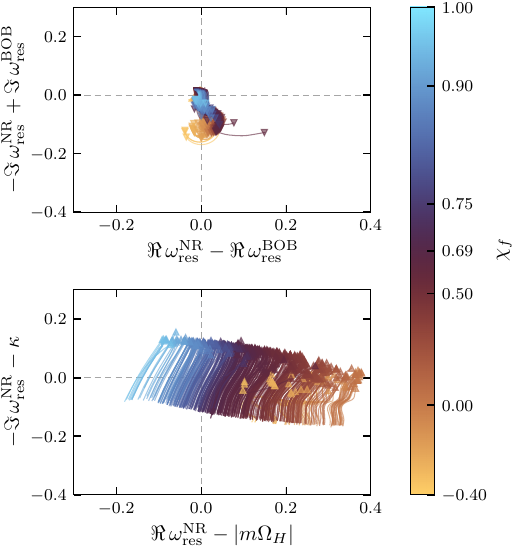}
    \caption{Top: Comparison of the difference between the NR residual non-QNM component and the finite N BOB analytical result over a $[-7.5,7.5]M$ window with respect to the unfiltered news peak. The arrow represents the difference at $+7.5M$.  Bottom: Comparison of the difference between the NR residual non-QNM component and the horizon surface gravity and frequency over the same window. Perfect agreement with the NR residual would lie on the intersection of the dashed grey lines. Appendix~\ref{app:bob_filtering} lists the details of the filtering process for this plot. The NR data consists of nearly 600 quasicircular and nonprecessing simulations.}
    \label{fig:bob_dw_catalog}
\end{figure}

We will refer to this result in our plots as the ``finite $N$'' result. Fig.~\ref{fig:bob_dw_catalog} compares the accuracy of this result over a time window of $[-7.5,7.5]M$ with respect to the peak of the unfiltered news waveform. Appendix~\ref{app:bob_filtering} provides the details on the NR and BOB filtering process. For comparison, we show the deviation of the NR non-QNM component from the horizon frequency and surface gravity. We note this figure compares the ``early time'' behavior of the filter. More recently, \cite{2026arXiv260823209S} has argued that the NR non-QNM component should asymptote to the horizon frequency at late times. Over this window, we find that the light ring based approach of BOB successfully models the NR non-QNM component across the quasicircular and non-precessing parameter space. Furthermore, this model matches the NR result across all remnant spins. The success of the light ring based approach of BOB in modeling this non-QNM component, over this time period, shows that a horizon calibration is not needed. Furthermore, we can also observe that a significant portion of the agreement between BOB and NR lies in the $s<0$ regime, corresponding to the UHP of Fig.~\ref{fig:pole_structure}. This part of the non-QNM component is a mix of the original unfiltered poles and the anti-causal poles introduced by the filter. Sufficiently before the peak, the non-QNM component of the filtered waveform consists largely of post-peak radiation that the filter has redistributed to earlier times. Unfiltered BOB models the post-peak radiation to high accuracy. Therefore, the filtered BOB is able to match the filtered NR result at times before the peak where unfiltered BOB has itself begun to break down, further showing that the filter is moving content from later to earlier times. We investigate the decomposition of the non-QNM component further in Sec.~\ref{sec:sub_decomposing_residual_main}.

\subsection{The Large $N$ limit}
Although we have obtained the finite $N$ result, BOB contains information from an infinite number of overtones in the eikonal limit \cite{Kankani:DWBOB}. In this section we take the large $N$ limit of our exact finite $N$ result. This approach necessitates several approximations, detailed in Appendix~\ref{app:BOB_derivation}, and we use the amplitude only for $s\leq0$.

In the large $N$ limit, the amplitude and frequency of the BOB residual are approximately given by
\begin{widetext}
\begin{align}
    A(t)_{\res,\bob, s<0} (t) &= 2A_p\sqrt{\frac{\sinh(\pi\lambda)}{\pi \lambda}} \xi(s) K_{i\lambda}(\xi(s)) \label{eq:BOB_DW_amp_largeN}\\
    \Omega_{\res,\bob} (t) &= \frac{1}{4}\Omega_0 + \frac{3}{4}\Omega_\qnm +\frac{1}{4}(\Omega_\qnm-\Omega_0)\tanh(s)
    \label{eq:BOB_DW_freq_largeN}
\end{align}
where $K_{i\lambda}$ is a modified Bessel function of the second kind.
Integrating Eq.~\eqref{eq:BOB_DW_freq_largeN} gives the phase while Eq.~\eqref{eq:BOB_DW_amp_largeN} provides the damping time $\tau_{\bob,\res}$ 
\begin{align}
    \Phi_{\res,\bob} (t) &= \frac{1}{4}\bigg(3\Omega_\qnm + \Omega_0 \bigg)t + \frac{\Omega_\qnm-\Omega_0}{4} \tau\ln (\cosh(s)) + \Phi_0 \\
    \tau_{\bob,\res}^{-1} &= \frac{d}{dt} \ln(A_{\bob,\res}(t))
     = \tau^{-1}\bigg(1 - \xi(s) \frac{\Re[K_{1+i\lambda}(\xi(s))]}{K_{i\lambda}(\xi(s))}\bigg)
     \label{eq:BOB_large_N_phase_and_damping}
\end{align}
\end{widetext}

At $s=0$ we obtain a value of 
\begin{align}
    \Omega_{\res,\bob} &= \frac{1}{4}\bigg(3\Omega_\qnm + \Omega_0 \bigg)
\end{align}
Noting that $\Omega_0/4 \approx 0.034$ -- $0.054$ and $\Omega_\qnm \approx 0.95\Omega_{\text{LR}}$, this frequency is very similar to the best fit result of $\sim 0.73 \Omega_{\textrm{LR}}$ found in Sec.~\ref{sec:DWfreq}\footnote{We note that $s=0$ corresponds to $t=t_p^{\mathcal{N}_{22}}$, so this value is not evaluated at the same time as the markers in Fig.~\ref{fig:spin_dependencies}. Because $t_p^{\mathcal{N}_{22}} - t_p^{h_{22}}$ is generally only a few $M$, and the frequency is very slowly evolving, the frequency does not change significantly in this window.}. Fig.~\ref{fig:BOB_3cases} compares the finite N and large N result against the NR non-QNM component. The finite N result matches the NR non-QNM component from $\sim 15$ -- $20M$ before the peak to $\sim 5$ -- $20M$ after the peak, depending on the system parameters. As expected, while the large N result performs well before the peak, the amplitude quickly breaks down after the peak.

\subsection{Decomposing the Residual}\label{sec:sub_decomposing_residual_main}
We have obtained an exact analytical expression for the BOB non-QNM component in Eqs.~\eqref{eq:BOB_dw_finite_N_sl0} and \eqref{eq:BOB_dw_finite_N_sg0}, and can now show that the ``loud'' part of the signal, corresponding to $s<0$, is some mix of the unfiltered poles in the UHP and the anti-causal poles introduced by the filter. However, it is not immediately clear from the foregoing analysis how large an effect the filter is introducing onto the waveform as a function of time. To better understand the decomposition of the non-QNM component, we extend the \textit{fourth way} of understanding the rational filter in \cite{2026arXiv260829466H}, which treats the filter as a convolution in the time domain. This allows the non-QNM component to be decomposed into the unfiltered waveform and a ``filter effects'' term. In Appendix~\ref{app:BOB_decompose} we extend this technique to removing multiple QNMs and specialize it to the case of BOB. While the details of this derivation are presented in Appendix~\ref{app:BOB_decompose}, schematically we can think of the non-QNM component as 
\begin{equation}
    \mathcal{N}_{\res,\bob}(t) = \mathcal{N}_\bob(t) + \textrm{filter effects}
\end{equation}
where $\mathcal{N}_\bob (t)$ is the unfiltered BOB news waveform. Eq.~\eqref{eq:2term_removal_operator} gives the exact decomposition for the removal of two QNMs, and the result extends to any number of QNMs.
\begin{figure}
    \centering
    \includegraphics[width=\linewidth]{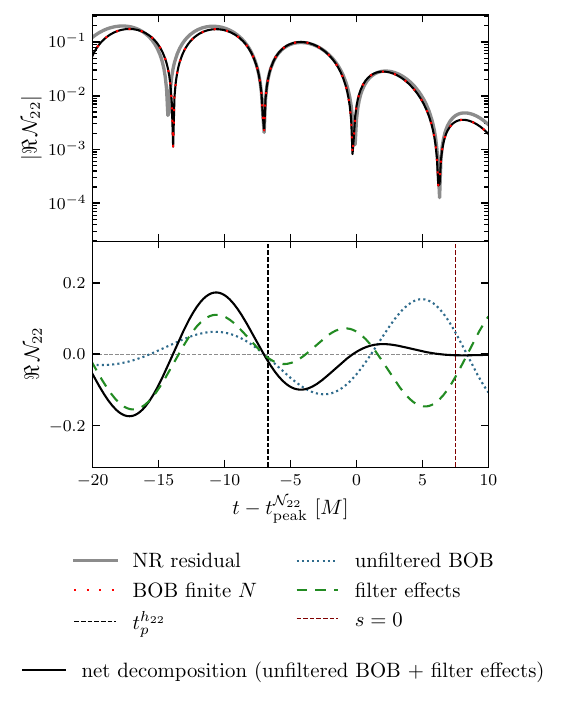}
    \caption{Top panel: The NR (solid grey) and BOB non-QNM component for \texttt{SXS:BBH:0305}. The BOB finite N result is calculated by solving Eq.~\eqref{eq:BOB_filtered_master} (dotted red). The BOB net decomposition result (solid black) is a time domain method for calculating the BOB non-QNM component that decomposes the residual into the unfiltered waveform (dotted blue) and filter effects (dashed green). Bottom panel: The unfiltered waveform and filter effects components of the net decomposition result. Details on the filtering process for this plot are provided in Appendix~\ref{app:bob_filtering}.}
    \label{fig:BOB_decomp_main}
\end{figure}
Fig.~\ref{fig:BOB_decomp_main} shows the decomposition of the BOB non-QNM component into an unfiltered part and a ``filter effects'' part. The exact details of the filtering process are provided in Appendix~\ref{app:bob_filtering}. Before $ t \approx -7M$, we find that the residual is dominated by the redistributed information from later times. From $t\approx -7M$ to $t\approx 0$, the unfiltered waveform appears to be the dominant portion of the residual. Afterwards, the two terms become approximately equal in strength and opposite in phase, leading to the damping of the residual. We interpret this phase as the filter redistributing QNM information to earlier times, revealing higher overtone information in the waveform. After $t\approx5M$, BOB starts showing disagreement with NR. This is expected as BOB only incorporates prograde linear QNM information. In Fig.~\ref{fig:BOB_decomp}, we repeat this procedure for \texttt{SXS:BBH:0178}, a large remnant spin case. This case shows clear differences in the decomposition in the time window between the strain and news peak. This suggests the near peak behavior of the filter is not universal, and must be understood individually for each waveform.

%% file: simplified_BOB.tex
\section{Simplifying BOB's Frequency}\label{app:BOB_simple_freq}
BOB connects the amplitude profile given in Eq.~\eqref{eq:BOB_amplitude} to the frequency by the approximate relationship \cite{baker2008mergers,mcwilliams2019analytical}
\begin{equation}
    |R\mathcal{N}_{lm}|^2 \approx 16\pi m^2 \frac{dJ^{\text{orb}}_\text{lm}}{d\Omega_{lm}}\Omega_{lm}\dot{\Omega}_{lm}. \label{eq:News_frequency}
\end{equation}
For the News, this results in the frequency equation:
\begin{align}
    \Omega_{\bob, \text{full}} (t) = \sqrt{\Omega^2_\qnm + \frac{\Omega^2_\qnm-\Omega^2_0}{2}\bigg[\tanh\bigg(\frac{t-t_p}{\tau}\bigg) - 1\bigg]}.
    \label{eq:BOB_news_frequency_evolution}
\end{align}
The waveform frequency $\omega$ is related to $\Omega$ by $\omega = m\Omega$, where $m$ is the gravitational wave mode number. $t_p$ is the time of the peak amplitude of the waveform. $\Omega_\qnm = \Re(\omega_{\qnm})/m$ gives the asymptotic limit of the frequency, where $\omega_{\qnm}$ is the true Kerr complex QNM. $\Omega_0$ is the initial frequency parameter, defined at $t=-\infty$, well before BOB becomes accurate.

Because of the nontrivial form of Eq.~\eqref{eq:BOB_news_frequency_evolution}, we are unable to perform an analytical Fourier transform of the resulting BOB time domain waveform. However, we observe that the general structure of the merger-ringdown frequency is governed primarily by a $\tanh(t/\tau)$ term. Furthermore, we have the physical asymptote of $\Omega_{\qnm}$ as $t\rightarrow \infty$ and the free parameter $\Omega_0$ at $t =-\infty$. Therefore, we propose a simpler form for the News frequency:
\begin{equation}
    \Omega_\bob = \frac{\Omega_{\qnm} + \Omega_0}{2} + \frac{\Omega_\qnm-\Omega_0}{2}\tanh\bigg(\frac{t-t_p}{\tau}\bigg).
     \label{eq:app_bob_freq_simple}
\end{equation}
Because the fit for $\Omega_0$ in \cite{kankani2026analyzing} used Eq.~\eqref{eq:BOB_news_frequency_evolution}, we refit $\Omega_0$ using Eq.~\eqref{eq:app_bob_freq_simple}. We perform a least squares fit of Eq.~\eqref{eq:app_bob_freq_simple} against the unfiltered NR waveform frequency, in the window [$t_p^{\mathcal{N}_{22}},t_p^{\mathcal{N}_{22}} +75M $], across nearly 600 quasicircular and nonprecessing SXS simulations. Across these simulations we obtain the fit:
\begin{equation}
    \Omega_0 = 0.148271 + 0.024652\chi_f + 0.038459 \chi_f^3
\end{equation}

\begin{figure*}
    \includegraphics[width=\textwidth]{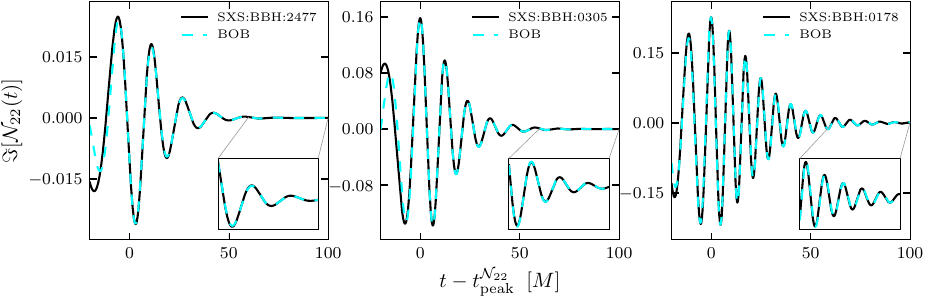}
    \caption{Comparison of the simplified BOB model with three NR waveforms for the imaginary part of the unfiltered News. \texttt{SXS:BBH:2477} is a high mass ratio ($q=15$) and low remnant spin ($\chi_f = 0.19$) system. \texttt{SXS:BBH:0305} is a near equal mass ($q=1.22$) system with moderate remnant spin ($\chi_f = 0.69$). \texttt{SXS:BBH:0178} is an equal mass system with an extremely large remnant spin ($\chi_f = 0.95$).}
    \label{fig:BOB_simple_3figs}
\end{figure*}

Fig.~\ref{fig:BOB_simple_3figs} compares BOB using the the simplified frequency model, given in Eq.~\eqref{eq:app_bob_freq_simple}, to three SXS simulations spanning different regions of the parameter space. \texttt{SXS:BBH:2477} is a high mass ratio ($q=15$) and low remnant spin ($\chi_f = 0.19$) system. \texttt{SXS:BBH:0305} is a near equal mass ($q=1.22$) system with moderate remnant spin ($\chi_f = 0.69$). This system has parameters similar to those of the first gravitational wave detection, GW150914, and has been extensively used in rational filter studies \cite{ma2022quasinormal,Oshita:2025qmn,2026arXiv260823209S,Kankani:DWBOB,Kankani:2026kst}. \texttt{SXS:BBH:0178} is an equal mass system with an extremely large remnant spin ($\chi_f = 0.95$). As seen in Fig.~\ref{fig:BOB_simple_3figs}, the simplified BOB frequency still results in a highly accurate merger-ringdown model from the peak onwards.

Integrating Eq.~\eqref{eq:app_bob_freq_simple} in the time domain gives us the phase evolution,
\begin{equation}
    \Phi_\bob = \frac{\Omega_\qnm + \Omega_0}{2}t + \frac{\Omega_\qnm - \Omega_0}{2}\tau \ln (\cosh(s)) +\Phi_0,
    \label{eq:BOB_news_phase_evolution}
\end{equation}
where $\Phi_0$ is an arbitrary phase offset. This gives us the simplified time domain waveform 
\begin{align}
    \mathcal{N}_\bob (t) &= A_p\operatorname{sech}(s) e^{-i\lambda\ln(\cosh(s))} 
    e^{-im\bar{\Omega}{(s\tau+t_p)}}.
\end{align}
Defining $\lambda = (m\tau/2)(\Omega_\qnm - \Omega_0)$ and $\bar{\Omega} = (\Omega_\qnm+\Omega_0)/2$, we can rewrite the waveform as
\begin{equation}
    \mathcal{N}_\bob (t) = A_p\text{sech}(s)^{1+i\lambda}e^{-im\bar{\Omega}(s\tau+t_p)}.
\end{equation}
The Fourier transform is then given by:
\begin{equation}
    \mathcal{N}_\bob (\omega) = A_p\tau e^{i\nu t_p/ \tau}\int_{-\infty}^{\infty} \text{sech}(s)^{1+i\lambda}e^{i\nu s}\, ds,
\end{equation}
where $ \nu = (\omega-m\bar{\Omega})\tau$. Using $x=e^{2s}$ so that $\text{sech}(s) = 2\sqrt{x}/(1+x)$, we can rewrite the integral as: 
\begin{equation}
    \mathcal{N}_\bob (\omega) = A_p\tau e^{i\nu t_p/ \tau}\int_0^{\infty} \bigg(\frac{2\sqrt{x}}{1+x}\bigg)^{1+i\lambda} \frac{x^{\frac{i\nu}{2}}}{2x}dx.
\end{equation}
This integral can be solved exactly\footnote{The SymPy code is provided at \href{https://github.com/AnujKankani/BOB_analytical_rational_filter}{\faGithub}}, giving us Eq.~\eqref{eq:BOB_fd}.

%% file: BOB_appendix.tex
\section{Obtaining the BOB residual} \label{app:BOB_derivation}
Because of the lengthy form of the equations involved in the derivation, we make the following substitutions:
\begin{align}
    &s = \frac{t-t_p}{\tau} \hspace{1.4cm}  \hspace{1.125cm}\lambda = \frac{m\tau}{2}(\Omega_\qnm - \Omega_0) \notag \\
    &\bar{\Omega} = \frac{\Omega_\qnm+\Omega_0}{2} \hspace{1.5cm}   \nu = (\omega-m\bar{\Omega})\tau  \notag\\
    &z_+(s) = e^{+is\lambda} \hspace{2cm} z_-(s) = e^{-is\lambda} \notag\\ \notag\\
    &\xi(s) = e^{s+\ln(2N)} \notag\\
    &\zeta(s) = A_pe^{-im\bar{\Omega}(t_p + s\tau)}2^{1+i\lambda} \notag\\
    & \kappa_1(s) = \,_2F_1(2+N+i\lambda,i\lambda-N,1+i\lambda,-e^{2s})  \notag\\
    &\kappa_2(s) = \,_2F_1(-N,2+N,1-i\lambda,-e^{2s})  \notag\\
    &\kappa_3(s) = \,_2F_1(2+N+i\lambda,2+N,3+2N,-e^{-2s}) \notag\\
    &\kappa_4(s,n) = \,_2F_1(1+i\lambda,n+1,n+2,-e^{-2s})
\end{align}
where $_2F_1(a,b,c;z)$ is the hypergeometric function given by
\begin{equation}
    _2F_1(a,b,c;z) =\sum_{n=0}^\infty \frac{(a)_n (b)_n}{(c)_n}\frac{z^n}{n!}.
\end{equation}
and $(x)_n = \Gamma(x+n)/\Gamma(x)$ is the Pochhammer symbol.
\subsection{Applying the filter to BOB}
Here, we detail how to solve Eq.~\eqref{eq:BOB_filtered_master} exactly to obtain Eqs.~\eqref{eq:BOB_dw_finite_N_sl0} and \eqref{eq:BOB_dw_finite_N_sg0}. We show how we take the large $N$ limit to obtain Eqs.~\eqref{eq:BOB_DW_amp_largeN}--\eqref{eq:BOB_large_N_phase_and_damping}. We provide detailed comparisons of our results to three specific SXS simulations. Parts of this calculation were performed using SymPy \cite{sympy} and the corresponding codes, as well as the key results, are provided at \href{https://github.com/AnujKankani/BOB_analytical_rational_filter}{\faGithub}
\begin{widetext}
\subsection{Pole Residues}
 Writing the eikonal filter in Eq.~\eqref{eq:eikonal_filter} in terms of $\nu$ and $ \lambda$ gives
\begin{equation}
    F_{p,\bob}(\nu) = \prod_{n=0}^{N} \frac{\nu - \lambda +i(2n+1)}{\nu - \lambda - i(2n+1)}.
    \label{eq:BOB_filter}
\end{equation}
The prograde BOB filter $F_{p,\bob}$ contains N+1 simple poles at $\nu = \lambda + i(2n+1)$. At any pole k, the filter can be written as
\begin{equation}
    F_{p,\bob}(\nu) = \frac{\nu - \lambda + i(2k+1)}{\nu - \lambda - i(2k+1)}\prod_{n\neq k}^{N} \frac{\nu - \lambda +i(2n+1)}{\nu - \lambda - i(2n+1)}.
\end{equation}
Fig.~\ref{fig:pole_structure} shows the pole structure after applying the filter to BOB. We calculate the residues of the BOB and filter poles as \footnote{The SymPy code is provided at \href{https://github.com/AnujKankani/BOB_analytical_rational_filter}{\faGithub}}

\begin{align}
\text{Res}_{\nu_k}[F_{p,\bob}] &= 2i(2k+1)\prod_{n\neq k}^{N} \frac{k+n+1}{k-n} = 2i (-1)^{N-k}\frac{(1+k+N)!}{(k!)^2(N-k)!}\\
\text{Res}_{\nu}[{\mathcal{N}_{\bob, s<0}(\omega)}] &= A_p\tau e^{(2n+1+i\lambda)\frac{-t_p}{\tau}}2^{i\lambda+1}(-1)^{n+3/2}\frac{\Gamma(1+i\lambda+n)}{\Gamma(1+n)\Gamma(1+i\lambda)}\\
\text{Res}_{\nu}[{\mathcal{N}_{\bob, s>0}(\omega)}] &= A_p\tau e^{(2n+1+i\lambda)\frac{t_p}{\tau}}2^{i\lambda+1}(-1)^{n+1/2}\frac{\Gamma(1+i\lambda+n)}{\Gamma(1+n)\Gamma(1+i\lambda)}
\end{align}
All the poles are simple, allowing us to easily apply the residue theorem to solve Eq.~\eqref{eq:BOB_filtered_master}. We use two contours, one closing in the upper half plane for $s<0$ and one closing in the lower half plane for $s>0$. The poles in the UHP are located at $\nu = \lambda + i(2n+1)$ and $\nu = -\lambda + i(2n+1)$. Applying the residue theorem for our contour in the UHP we obtain, 
\begin{align}
 \mathcal{N}_{\res,\bob, s<0} (t) = \zeta(s)\bigg[&\sum_{n}\bigg( z_+(s)e^{s(2n+1)} (-1)^{n}
 \frac{\Gamma(i\lambda + 2 + n+ N)\Gamma(i\lambda +n-N)}{\Gamma(i\lambda + 1 +n)\Gamma(1+n)\Gamma(i\lambda + 1)}\bigg)\notag\\
  - &\sum_k z_-(s)e^{s(2k+1)} (-1)^{N-k}   \frac{\Gamma(i\lambda - k)\Gamma(2+k+N)}{\Gamma(k+1)\Gamma(N-k+1)\Gamma(1+i\lambda)}\bigg], 
\end{align}
where we have used 
\begin{align}
    \prod_{m=0}^{N} \frac{i\lambda + 1 + n + m}{i\lambda + n - m} &= \frac{\Gamma(i\lambda + 2 + n+ N)\Gamma(i\lambda +n-N)}{\Gamma(i\lambda + 1 +n)^2}.
\end{align}
$\sum_n$ is over the BOB poles in the UHP and $\sum_k$ is over the filter poles in the UHP. We can simplify this expression using $_2F_1(a,b;c;z)$ hypergeometric functions:
\begin{align}
     \mathcal{N}_{\res,\bob, s<0} (t)  = -\zeta(s)e^s \bigg[-z_+(s)\kappa_1(s)\frac{\Gamma(2+N+i\lambda)\Gamma(i\lambda-N)}{\Gamma(1+i\lambda)^2}
    +  z_-(s)\kappa_2(s) (-1)^{N}\frac{\Gamma(i\lambda)\Gamma(2+N)}{\Gamma(1+N)\Gamma(1+i\lambda)}\bigg].
\end{align}
Closing in the LHP we obtain
\begin{align}
\mathcal{N}_{\res,\bob, s>0} (t) = \zeta(s)\bigg[&\sum_{n>N} z_-(s)e^{-s(2n+1)} (-1)^{n}\frac{\Gamma(1+n)^2\Gamma(1+i\lambda+n)}{\Gamma(n-N)\Gamma(2+n+N)\Gamma(1+n)\Gamma(1+i\lambda)} \bigg], 
\end{align}
where we have used
\begin{align}
    \prod_{m=0}^N \frac{n-m}{1+n+m} &= \frac{\Gamma(1+n)^2}{\Gamma(n-N)\Gamma(2+n+N)}.
\end{align}
For the LHP, only the $n>N$ poles remain after applying the filter. Shifting the summation such that $n=N+1+j$ we obtain
\begin{align}
    & \mathcal{N}_{\res,\bob, s>0} (t) = - \zeta(s)z_-(s) (-1)^{N} \bigg[\sum_{j=0} e^{-s(3+2N+2j)} (-1)^{j}\frac{\Gamma(2+N+j)\Gamma(i\lambda+2+N+j)}{\Gamma(1+j)\Gamma(3+2N+j)\Gamma(1+i\lambda)} \bigg]. 
\end{align}
We can also rewrite this equation in terms of a $_2F_1(a,b;c;z)$ hypergeometric function: 
\begin{align}
\mathcal{N}_{\res,\bob, s>0} (t) = -& \zeta(s)z_-(s) (-1)^{N} e^{-s(3+2N)} \frac{\Gamma(2+N)\Gamma(i\lambda + 2 +N)}{\Gamma(3+2N)\Gamma(1+i\lambda)}\kappa_3(s).
\end{align}
We now have two different equations for our two contours. However, we can show these are two parts of the same function by using the following hypergeometric identity \cite[\href{https://dlmf.nist.gov/15.8.E2}{(15.8.2)},\href{https://dlmf.nist.gov/15.2.E2}{(15.2.2)}]{NIST:DLMF} 
\begin{align}
{}_2F_1(a,b;c;z) 
= \frac{\Gamma(c)\,\Gamma(b-a)}{\Gamma(b)\,\Gamma(c-a)}\,(-z)^{-a}\,
  {}_2F_1\!\left(a,\,1-c+a;\,1-b+a;\,\tfrac{1}{z}\right)
+ \frac{\Gamma(c)\,\Gamma(a-b)}{\Gamma(a)\,\Gamma(c-b)}\,(-z)^{-b}\,
  {}_2F_1\!\left(b,\,1-c+b;\,1-a+b;\,\tfrac{1}{z}\right).
\end{align}
Using this identity, we can show that
\begin{align}
\centering
    e^{-s(3+2N)} \frac{\Gamma(2+N)\Gamma(i\lambda + 2 +N)}{\Gamma(3+2N)\Gamma(1+i\lambda)} z_-(s) \kappa_3(s) = \frac{\Gamma(2+N +i\lambda)\Gamma(-i\lambda)}{\Gamma(1+N-i\lambda)\Gamma(1+ i\lambda)}e^{s}z_+(s) \kappa_1(s) + \frac{\Gamma(2+N)\Gamma(i\lambda)}{\Gamma(1+i\lambda)\Gamma(1+N)}e^{s}z_-(s)\kappa_2(s). 
\end{align}
Noting that 
\begin{equation}
    \frac{(-i)\Gamma(i\lambda-N)}{\Gamma(1+i\lambda)} = \frac{(-1)^{5/2+N}\Gamma(-i\lambda)}{\Gamma(1+N-i\lambda)},
\end{equation}
we see that our $s<0$ and $s>0$ solutions to Eq.~\eqref{eq:BOB_filtered_master}, Eqs.~\eqref{eq:BOB_dw_finite_N_sl0} and \eqref{eq:BOB_dw_finite_N_sg0}, are two parts of the same continuous function.
\end{widetext}

\subsection{The Large $N$ limit}
While Eqs.~\eqref{eq:BOB_dw_finite_N_sl0} and ~\eqref{eq:BOB_dw_finite_N_sg0} represent the exact result from the application of the eikonal rational filter to the BOB waveform, obtaining the large $N$ limit requires certain approximations that we detail here.
\subsubsection{Obtaining the Phase}
We rewrite the hypergeometric function in Eq.~\eqref{eq:BOB_dw_finite_N_sg0} in its integral form \cite[\href{https://dlmf.nist.gov/15.6.E1}{(15.6.1)},\href{https://dlmf.nist.gov/15.2.E2}{(15.2.2)}]{NIST:DLMF} 
\begin{align}
    \kappa_3(s) = \frac{\Gamma(2N+3)}{\Gamma(N+2)\Gamma(N+1)}\int_0^1 \frac{t^{N+1}(1-t)^N}{(1+e^{-2s}t)^{2+N+i\lambda}}dt.
    \label{eq:hypergeometric_integral_form}
\end{align}
This can be rewritten as
\begin{equation}
    \int_0^1 \bigg(\frac{t(1-t)}{(1+e^{-2s}t)}\bigg)^N t(1+e^{-2s}t)^{-(2+i\lambda)}dt = \int_0^1 g(t) e^{NF(t)}dt,
\end{equation}
where $F(t) = \ln (t) + \ln (1-t) - \ln (1+e^{-2s}t)$ and $g(t) = t(1+e^{-2s}t)^{-(2+i\lambda)}$.
Laplace's method, which approximates the peak of the integrand as a Gaussian of width $\sigma = 1/\sqrt{N|F^{''}(t_0)|}$, where $t_0$ is the location of the maximum of $F(t)$, becomes an obvious choice to solve this integral. For our integral,
\begin{align}
&t_0 = e^{2s}(\sqrt{1+e^{-2s}}-1) \\
&e^{NF(t_0)} = (t_0)^{2N}\\
&F^{''}(t_0) = -2/(t_0(1-t_0)).
\end{align}
However, care must be taken when $s$ is sufficiently before the peak to ensure that the Gaussian does not get cut off by the lower bound of the integral. 
For valid $s$, the integral can be approximated through Laplace's method
\begin{equation}
    \int_0^1 g(t)e^{NF(t)}dt \approx \sqrt{\frac{2\pi}{N|F^{''}(t_0)|}}g(t_0)e^{NF(t_0)}.
    \label{eq:laplace_method}
\end{equation}
However, we can recognize that all of the \textit{phase information} carried by the hypergeometric function in Eq.~\eqref{eq:BOB_dw_finite_N_sg0} is contained at the peak of the Gaussian $g(t_0)$. Therefore, this method can still give us the full phase information carried by our hypergeometric function. 
The asymptotics of the gamma functions appearing in Eqs.~\eqref{eq:BOB_dw_finite_N_sg0} and \eqref{eq:hypergeometric_integral_form} are independent of Laplace's method and are therefore unaffected by this issue and can be obtained using \cite[\href{https://dlmf.nist.gov/5.5.E5}{(5.5.5)},\href{https://dlmf.nist.gov/5.11.E12}{(5.11.12)}]{NIST:DLMF}. We obtain the following \textit{incomplete} result for the rational filter residual applied to BOB. 
\begin{widetext}
\begin{align}
\mathcal{N}_{\res,\bob}(t) \sim -\zeta(s)z_-(s)e^{-3s} N^{i\lambda}\sqrt{\pi(N+1)(1-t_0)}(t_0)^{\frac{4N+3}{2}}(1+e^{-2s}t_0)^{-(2+i\lambda)} (-e^{-2s})^N \frac{1}{\Gamma(1+i\lambda)}. 
\end{align}
This result contains only the correct phase information. Therefore, we can obtain the frequency 
\begin{align}
    \omega(t) = -\Im\bigg[\frac{1}{\mathcal{N}_{\res,\bob}(t)}\frac{d(\mathcal{N}_{\res,\bob}(t))}{dt}\bigg]
    = m\bar{\Omega} - \frac{1}{\tau}\frac{d}{ds}\arg\bigg[z_-(s)e^{-s(2N+3)}\kappa_3(s)\bigg].
\end{align}
Solving this gives us the leading-order, BOB-predicted non-QNM frequency as $N\rightarrow\infty$
\begin{align}
    \Omega_{\res,\bob} (s) = \frac{1}{4}\Omega_0 + \frac{3}{4}\Omega_\qnm +\frac{1}{4}(\Omega_\qnm-\Omega_0)\tanh (s).
\end{align}
\end{widetext}
\subsubsection{Obtaining the Amplitude}
To obtain the amplitude for $s<0$, we approximate our $_2F_1$ hypergeometric functions in Eq.~\eqref{eq:BOB_dw_finite_N_sl0} as modified Bessel functions of the first kind. We define $x = 2Ne^s$ so that $-e^{2s} = -x^2/(4N^2)$. Using $N\rightarrow\infty$, $(N)_n \sim N^n$, we make the approximations:
\begin{align}
    &(2+N+i\lambda)_n(i\lambda-N)_n \label{eq:amp_poch1}\\&\sim (-N^2-2N-\lambda^2+2i\lambda)^n \approx (-N^{2})^{n} \notag \\
    &(-N)_n(2+N)_n \sim (-N^2-2N)^n\approx (-N^{2})^{n} \label{eq:amp_poch2}
\end{align}
We note that $\lambda \sim \mathcal{O}(1)$, but dropping the $2i\lambda$ factor in Eq.~\eqref{eq:amp_poch1} spoils the phase information. The amplitude, however, can still be obtained. With the approximations above, in the large $N$ limit, we can turn our $_2F_1$ hypergeometric functions into $_0F_1$ hypergeometric functions
\begin{align}
    &\kappa_1(s) \rightarrow \, _0F_1(1+i\lambda;x^2/4) \notag \\
    &\kappa_2(s) \rightarrow \, _0F_1(1-i\lambda;x^2/4).
\end{align}
We can then convert our $_0F_1$ hypergeometric functions into modified Bessel functions of the first kind using \cite[\href{https://dlmf.nist.gov/10.39.E9}{(10.39.9)}]{NIST:DLMF} 
\begin{equation}
    _0F_1(1+\nu;z^2/4) = \frac{I_\nu (z) \Gamma(\nu +1)}{(\frac{1}{2}z)^\nu}
\end{equation}
Therefore, the two hypergeometric functions appearing in Eq.~\eqref{eq:BOB_dw_finite_N_sl0} can be approximated in the large $N$ limit as
\begin{align}
    & \kappa_1(s) \rightarrow \frac{I_{i\lambda}(x)\Gamma(1+i\lambda)}{(x/2)^{i\lambda}} \notag \\
    & \kappa_2(s) \rightarrow \frac{I_{-i\lambda}(x)\Gamma(1-i\lambda)}{(x/2)^{-i\lambda}} ,
    \label{eq:hyper_bessel_identity}
\end{align}
where $I_\nu(z)$ is a modified Bessel function of the first kind with complex argument $\nu$ and $z>0$ \cite[\href{https://dlmf.nist.gov/10.39.E9}{(10.39.9)}]{NIST:DLMF}.
By rewriting
\begin{equation}
    \Gamma(i\lambda-N) = \frac{-i\pi}{(-1)^N \sinh(\pi\lambda)\Gamma(1+N-i\lambda)}
\end{equation}
\cite[\href{https://dlmf.nist.gov/5.5.E3}{(5.5.3)},\href{https://dlmf.nist.gov/4.28.E8}{(4.28.8)}]{NIST:DLMF} and using the asymptotic behavior $\Gamma(z+a)/\Gamma(z+b) \sim z^{a-b}$ \cite[\href{https://dlmf.nist.gov/5.11.E12}{(5.11.12)}]{NIST:DLMF} we obtain
\begin{align}
    & \frac{\Gamma(2+N+i\lambda)\Gamma(i\lambda-N)}{\Gamma(1+i\lambda)^2} \sim \frac{(-1)^N(-i\pi) N^{1+2i\lambda}}{\sinh(\pi\lambda)\Gamma(1+i\lambda)^2} \\
    & \frac{\Gamma(2+N)\Gamma(i\lambda)}{\Gamma(1+i\lambda)\Gamma(1+N)} \sim \frac{-i\pi N}{\sinh(\pi \lambda)\Gamma(1+i\lambda)\Gamma(1-i\lambda)}.
\end{align}
Combining our approximations allows us to obtain an \textit{incomplete} result for Eq.~\eqref{eq:BOB_dw_finite_N_sl0} in the large $N$ limit for $s<0$ 
\begin{equation}
    \mathcal{N}_{\res,\bob, s<0} (t) \sim -\zeta(s)e^{s}  \frac{2 (-1)^N}{ \Gamma(1+i\lambda)}N^{1 + i\lambda} K_{i\lambda}(2Ne^s)
\end{equation}
where $K_\nu(z)$ is the modified Bessel function of the second kind \cite[\href{https://dlmf.nist.gov/10.27.E4}{(10.27.4)}]{NIST:DLMF} with complex argument $\nu$ and $z>0$.
\begin{equation}
    K_\nu(z) = \frac{\pi}{2\sin(\pi\nu)}(I_{-\nu}(z) - I_\nu(z))
\end{equation}

However, this equation contains only the correct \textit{amplitude} information. Using the identity \cite[\href{https://dlmf.nist.gov/5.4.E3}{(5.4.3)}]{NIST:DLMF} $|\Gamma(i\lambda)| = \sqrt{\pi/(\lambda\sinh(\pi \lambda))}$, we can read off the amplitude as 
\begin{equation}
    A_{\res,\bob, s<0} (t) = 2A_p\sqrt{\frac{\sinh(\pi\lambda)}{\pi \lambda}} \xi(s) K_{i\lambda}(\xi(s)),
\end{equation}
where we have rewritten $2Ne^{s}$ as $\xi(s)$ to make it clear that while the result does depend on $N$, it enters only as an effective time shift. 

The imaginary part of the complex frequency $\Omega_I$ is then given by 
\begin{align}
    \Omega_I &= \frac{d}{dt} \ln(A(t)) \notag \\ &= \frac{1}{\tau}\bigg(1+i\lambda - \xi(s) \frac{K_{1+i\lambda}(\xi(s))}{K_{i\lambda}(\xi(s))}\bigg) \notag \\
     &= \frac{1}{\tau}\bigg(1 - \xi(s) \frac{\Re[K_{1+i\lambda}(\xi(s))]}{K_{i\lambda}(\xi(s))}\bigg)
\end{align}

\begin{figure*}
    \centering
    \includegraphics[width=\linewidth, height=0.9\textheight]{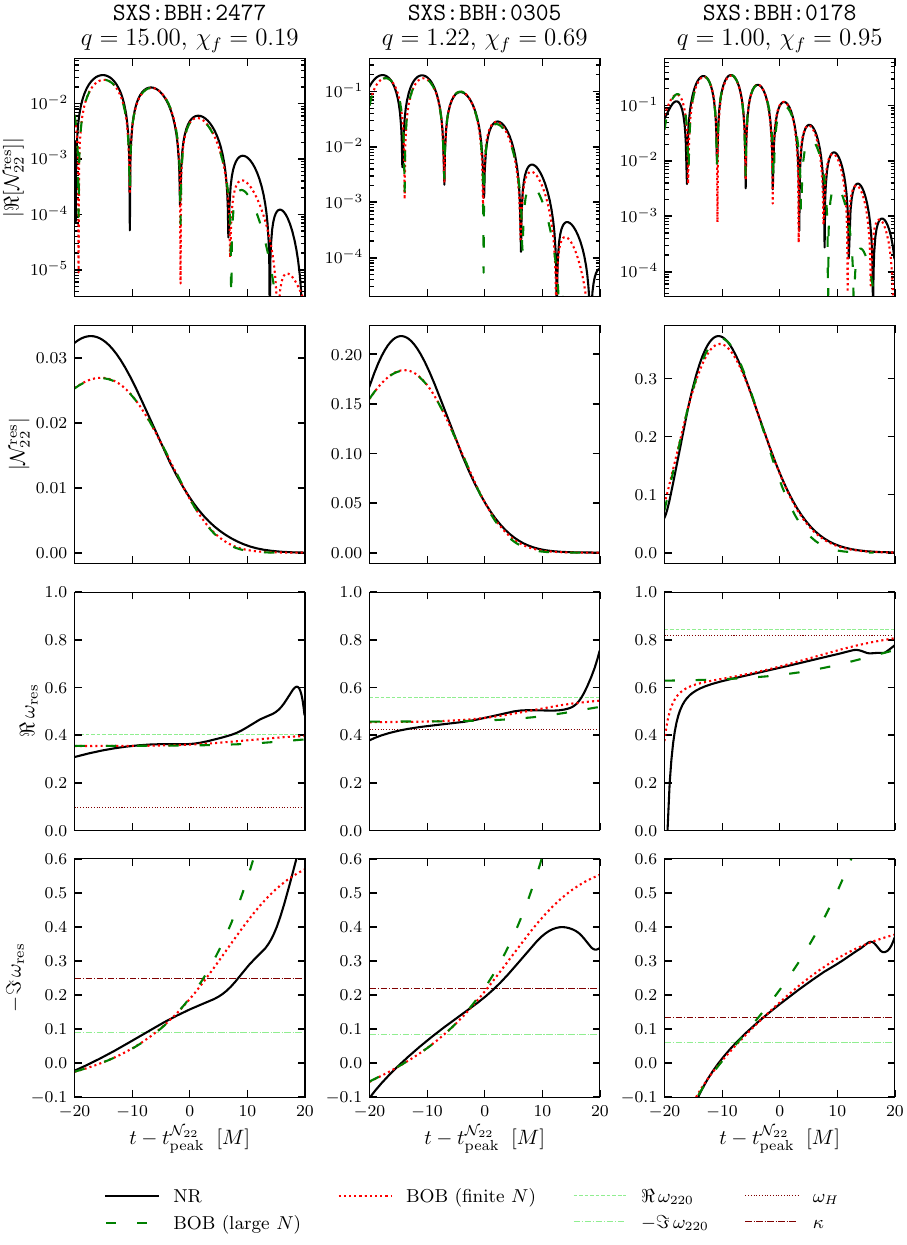}
    \caption{Comparison of the NR non-QNM component with the BOB analytical results. The finite N result is an exact result obtained by solving Eq.~\eqref{eq:BOB_filtered_master}. The large N result requires several approximations, resulting in a degradation of accuracy, particularly for the amplitude. We plot the absolute value of the real part of the non-QNM component (first row), the amplitude (second row), the frequency (third row) and the inverse damping time (bottom row). Details on the filtering process for this plot are provided in Appendix~\ref{app:bob_filtering}.}
    \label{fig:BOB_3cases}
\end{figure*}

%% file: BOB2_appendix.tex
\section{Decomposing the Residual}\label{app:BOB_decompose}
Here we extend to BOB the \textit{fourth way} of understanding the rational filter in \cite{2026arXiv260829466H}. This method treats the filter as a convolution in time, 
\begin{equation}
    \Psi(t) \rightarrow\hat{\Psi}(t) \equiv \frac{1}{\sqrt{2\pi}} \int_{-\infty}^{\infty}F(\sigma)\Psi(t-\sigma)\,d\sigma
\end{equation}
where
\begin{equation}
    F(\sigma) = \frac{1}{\sqrt{2\pi}}\int_{-\infty}^{\infty} \mathcal{F}(\omega)e^{-i\omega \sigma}\,d\omega.
\end{equation}
As shown in \cite{2026arXiv260829466H}, one can obtain the result
\begin{equation}
    \hat{\Psi}(t) = \Psi(t) + a_k\mathcal{D}(t),
\end{equation}
where $a_k = 2\Im(\omega_{220})$ and
\begin{equation}
    \mathcal{D}(t) = \int_0^\infty e^{i\omega_{220}^*\sigma}\Psi(t+\sigma)d\sigma.
    \label{eq:operator}
\end{equation}
To remove additional QNMs, we apply $\mathcal{D}$ repeatedly. Writing $\mathcal{D}_0$ for the first application, the second application $\mathcal{D}_1$ is
\begin{align}
    \mathcal{D}_1 &= \int_0^\infty\mathcal{D}_0(t+\sigma')e^{i{\omega_{221}^*}\sigma'}d\sigma' \notag\\
     &= \int_0^\infty e^{i{\omega_{221}^*}\sigma'}d\sigma'\int_0^\infty \Psi(t+\sigma+\sigma')e^{i\omega_{220}^*\sigma}\,d\sigma.
\end{align}
Letting $x = \sigma+\sigma'$ allows us to rewrite this as 
\begin{align}
    \mathcal{D}_1 = \int_0^\infty e^{i{\omega_{221}^*}\sigma'}d\sigma'&\int_{\sigma'}^\infty \Psi(t+x)e^{i\omega_{220}^*(x-\sigma')}\,dx \\
    = \int_0^\infty \Psi(t+x) dx&\int_0^x e^{i\omega_{221}^*\sigma'}e^{i\omega_{220}^*(x-\sigma')}\,d\sigma' \notag\\
    = \frac{1}{i\omega_{221}^* - i\omega_{220}^*}\bigg[&\int_0^\infty \Psi(t+x)e^{i\omega_{221}^*x}\,dx \notag\\ - &\int_0^\infty \Psi(t+x)e^{i\omega_{220}^*x}\,dx \bigg].\notag
\end{align}
Therefore, the subsequent application of two $\mathcal{D}$ operators $\mathcal{D}_1[\mathcal{D}_0(t)]$ can be written as 
\begin{equation}
    \mathcal{D}_1[\mathcal{D}_0(t)] = \frac{\mathcal{D}_0 - \mathcal{D}_1}{i(\omega_{220}^* - \omega_{221}^*)}.
\end{equation}
Repeating this process gives the relation for an arbitrary number of $\mathcal{D}$ operations
\begin{equation}
    \mathcal{D}_{N}\cdots\mathcal{D}_{1}\mathcal{D}_{0}(t) = \sum_{m=0}^N\frac{\mathcal{D}_m}{\prod\limits_{n=0, n\neq m}^N (i\omega_{22m}^*-i\omega_{22n}^*)}.
\end{equation}
We can then write the net filtered waveform as
\begin{equation}
    \Psi_\res (t) = \bigg[\prod\limits_{n=0}^N (1 +a_n\mathcal{D}_n)\bigg] \Psi(t).
\end{equation}
As an example, removing two QNMs gives
\begin{align}
    \Psi_\res(t) = (1&+a_0\mathcal{D}_0+a_1\mathcal{D}_1 + a_0a_1\mathcal{D}_0\mathcal{D}_1)\Psi(t) \notag \\
    = \Psi(t) &+ a_0\mathcal{D}_0\Psi(t) + a_1\mathcal{D}_1\Psi(t) \notag \\ &+ a_0a_1\frac{\mathcal{D}_1\Psi(t)}{i(\omega_{221}^* - \omega_{220}^*)}-a_0a_1\frac{\mathcal{D}_0\Psi(t)}{i(\omega_{221}^* - \omega_{220}^*)},\notag
\end{align}
Rearranging the terms, we obtain: 
\begin{align}
   \Psi_\res(t)  = \Psi(t) &+ a_0\bigg(1-\frac{a_1}{i(\omega_{221}^* - \omega_{220}^*)}\bigg)\mathcal{D}_0[\Psi(t)]\notag \\ &+a_1\bigg(1+\frac{a_0}{i(\omega_{221}^* - \omega_{220}^*)}\bigg)\mathcal{D}_1[\Psi(t)],
   \label{eq:2term_removal_operator}
\end{align}
where $a_n = 2\Im(\omega_{22n})$. This can be trivially applied to any number $N$ of QNMs. The net result is a first term, $\Psi(t)$, depending only on the unfiltered waveform, plus a sequence of further terms, each obtained by applying Eq.~\eqref{eq:operator} to the unfiltered waveform. This route has the significant advantage of allowing us to determine at any given time how much of the non-QNM component comes from the original waveform and how much from the filter. While so far this derivation works for any generic waveform, $\Psi(t)$, we will now specialize to the case of the eikonal filter used for BOB.
\subsection{Specializing to BOB}
Switching to the eikonal filter, Eq.~\eqref{eq:eikonal_filter}, so that we can apply this technique to BOB, we set $\omega_{22n} = m\Omega_\qnm - i(2n+1)/\tau$. Setting $u=\sigma/\tau$ allows us to rewrite $\mathcal{D}(t)$ as 
\begin{equation}
    \mathcal{D}_{\bob,\,n}(t) = \int_0^\infty A_p \text{sech}(s+u)^{1+i\lambda}e^{-im\bar{\Omega}t}e^{u(i\lambda-(2n+1))}\tau \,du,
\end{equation}
where we have used the substitutions defined at the beginning of Appendix~\ref{app:BOB_derivation} and $n$ is the index number of the QNM. 
Rewriting the hyperbolic secant in terms of exponentials allows us to write
\begin{align}
    \mathcal{D}_{\bob,\,n}(t) = &\tau \zeta(s) e^{-s}z_-(s)\notag\\&\int_0^\infty(1+e^{-2(s+u)})^{-(1+i\lambda)}e^{-2u(n+1)}\,du
\end{align}
We can recast this in a simpler form, 
\begin{equation}
    \mathcal{D}_{\bob,\,n}(t) = \tau \zeta(s)z_-(s)e^{-s}\frac{1}{2}\int_0^1 f^n(1+e^{-2s}f)^{-(1+i\lambda)}\,df ,
\end{equation}
where $f=e^{-2u}$. This allows us to recognize the integral form of the $_2F_1$ hypergeometric function \cite[\href{https://dlmf.nist.gov/15.6.E1}{(15.6.1)}]{NIST:DLMF}. $\mathcal{D}_\bob(t)$ therefore simplifies to
\begin{equation}
    \mathcal{D}_{\bob,\,n}(t) = \frac{\tau\zeta(s)z_-(s)e^{-s}}{2(n+1)}\,_2F_1(1+i\lambda,n+1;n+2;-e^{-2s})
\end{equation}
which in terms of the unfiltered BOB news $\mathcal{N}_\bob(t)$ is 
\begin{equation}
    \mathcal{D}_{\bob,\,n}(t) = \mathcal{N}_\bob(t) \frac{\tau(1+e^{-2s})^{1+i\lambda}\kappa_4(s,n)}{2(n+1)}
    \label{eq:operator_BOB}
\end{equation}
where $\kappa_4(s,n) = \,_2F_1(1+i\lambda,n+1;n+2;-e^{-2s})$. Therefore, removing a single QNM from BOB gives us the result
\begin{equation}
    \mathcal{N}_{\res,\bob}(t) = \mathcal{N}_\bob(t) - \frac{2}{\tau}\mathcal{D}_\bob(t)
\end{equation}
where $\mathcal{N}_\bob(t)$ is the unfiltered BOB time domain news waveform. The removal of additional QNMs follows similar to Eq.~\eqref{eq:2term_removal_operator} except instead of having to solve the integral in Eq.~\eqref{eq:operator}, we get to evaluate the hypergeometric function $\kappa_4$.
\begin{figure}
    \centering
    \includegraphics[width=\linewidth]{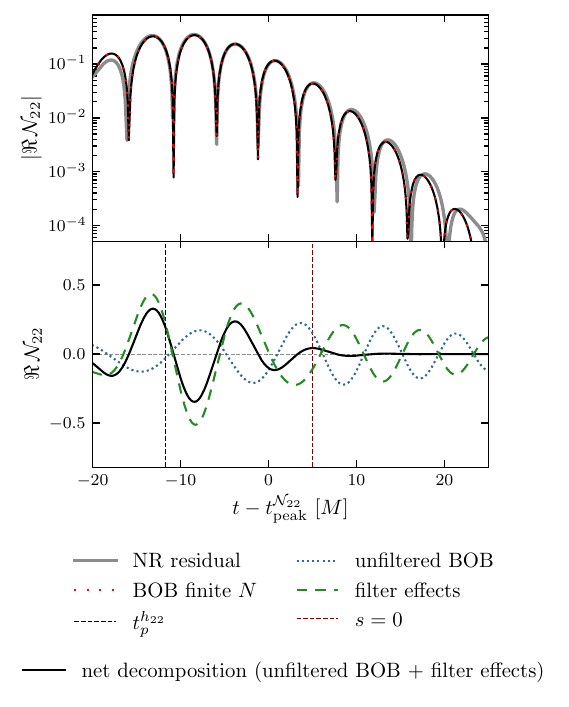}
    \caption{Top panel: The NR (solid grey) and BOB non-QNM component for \texttt{SXS:BBH:0178}. The BOB finite N result is calculated by solving Eq.~\eqref{eq:BOB_filtered_master} (dotted red). The BOB net decomposition result (solid black) is a time domain method for calculating the BOB non-QNM component that decomposes the residual into the unfiltered waveform (dotted blue) and filter effects (dashed green). Bottom panel: The unfiltered waveform and filter effects components of the net decomposition result. Details on the filtering process for this plot are provided in Appendix~\ref{app:bob_filtering}.}
    \label{fig:BOB_decomp}
\end{figure}
\section{BOB filtering} \label{app:bob_filtering}
Because the non-QNM component can be sensitive to the exact choice of QNMs removed, here we detail the specific filtering choices used in Sec.~\ref{sec:BOB} and Appendix~\ref{app:BOB_derivation}. A common filtering choice for the $(2,2)$ mode is the removal of 7 prograde QNMs, 2 retrograde QNMs and the $(3,2,0)$ mode \cite{Oshita:2025qmn,Kankani:2026kst}. However, \cite{2026arXiv260823209S} pointed out the conflict between higher overtone poles and the Matsubara poles at higher remnant spins. Because of this, \cite{2026arXiv260823209S} only removed modes up to $n=2$. They also removed all of the $(l\leq6,m=2,n=0,1,2)$ modes from the $(2,2)$ mode. While mode mixing from the $(3,2,0)$ mode is expected in the $(2,2)$ mode, and \cite{ma2022quasinormal} explicitly found evidence for this mode in \texttt{SXS:BBH:0305} using rational filters, it is unclear to what extent removing a large number of $(3\leq l\leq 6,m=2,n=0,1,2)$ modes is necessary. 

Further complicating the issue, BOB does not suffer from mode mixing and incorporates only prograde QNM information in its amplitude evolution, Eq.~\eqref{eq:BOB_amplitude}. As a reasonable choice, recognizing the conflict between higher overtone poles and the Matsubara poles and also wanting to minimize the differences in the filter applied to BOB and NR, we remove the first three $(l=2,m=2)$ prograde and retrograde QNMs as well as the $(l=3,m=2,n=0)$ mode from NR when comparing to BOB. From BOB, we remove only the first three $(l=2,m=2,n)$ prograde modes. We stress that our findings relating to the lack of connection between the non-QNM component and horizon information are insensitive to any reasonable choice of QNMs. Sec.~\ref{sec:DWfreq} uses the exact filtering choice   of \cite{2026arXiv260823209S} and \cite{Kankani:2026kst} used the 7 prograde, 2 retrograde and $(3,2,0)$ mode filtering choice of \cite{Oshita:2025qmn}.

Because we remove a different number of QNMs from BOB and NR in Sec.~\ref{sec:BOB} and Appendix~\ref{app:BOB_derivation}, the filters will apply a different time shift in each case. For the NR non-QNM component, we align with the unfiltered waveform $1000M$ before the peak. For BOB, because it is purely a merger-ringdown model, we perform an overall time and phase shift by minimizing the amplitude and phase residual against the filtered NR waveform over a time window of $[-10,0]M$ with respect to the peak of the unfiltered waveform.